\documentstyle[11pt,aaspp4,tighten]{article}

\def\civ{C\,{\sc iv}}
\def\mg2{Mg\,{\sc ii}}
\def\mabs{$M_{abs}$}
\def\prad{$P_{rad}$}
\def\ks{$K_s$}
\def\rks{$r$$-$$K_s$}
\def\hks{$H$$-$$K_s$}
\def\jks{$J$$-$$K_s$}
\def\jk{$J$$-$$K$}
\def\jKuk{$J$$-$$K_{UKIRT}$}
\def\etal{{\it et\,al.}}
\def\ebv{$E$($B$$-$$V$)}
\def\ho{$H_0$}
\def\qo{$q_0$}
\def\q0{$q_0$}
\def\h0{$H_0$}
\def\hm1{$h^{-1}$}
\def\hn1{$h_{75}^{-1}$}
\def\n05{$N_{0.5}$}
\def\no5{$N_{0.5}$}

\def\bgq{$B_{gq}$}
\def\z{$z$}
\def\h{$h$}
\def\k{$K$}

\def\Kuk{$K_{UKIRT}$}
\def\K{$K$}

\def\rK{$r$$-$$K$}
\def\rk{$r$$-$$K$}
\def\Kz{$K$$-$$z$}
\def\th{$\theta$}
\def\sig{$\sigma$}

\def\kmsm{km~s$^{-1}$~Mpc$^{-1}$}

\lefthead{HALL AND GREEN}
\righthead{RLQ ENVIRONS II. IMAGING RESULTS}
\slugcomment{Accepted to The Astrophysical Journal on June 9, 1998}

\begin{document}

\title{AN OPTICAL/NEAR-INFRARED STUDY OF RADIO-LOUD QUASAR ENVIRONMENTS II. IMAGING RESULTS}

\author{Patrick B. Hall
\footnote{Visiting Student, Kitt Peak National Observatory, National Optical Astronomy Observatories, operated by AURA Inc., under contract with the National Science Foundation.}
\footnote{Current address:  Department of Astronomy, University of Toronto, 60 St. George Street, Toronto, Ontario, Canada M5S~3H8}}
\affil{Steward Observatory, The University of Arizona, Tucson, Arizona 85721 \\
Electronic Mail: hall@astro.utoronto.ca}
\author{Richard F. Green}
\affil{National Optical Astronomy Observatories, Tucson, Arizona 85726-6732\\
Electronic Mail: rgreen@noao.edu}

\begin{abstract}	\label{begin}

We have previously reported a significant excess of $K$$\gtrsim$19 galaxies
in the fields of a sample of 31 $z$=1--2 quasars (\cite{hgc98}).
Here we examine the properties of this excess galaxy population
using optical and near-IR imaging.  

The excess occurs on two spatial scales.
One component lies at $\theta$$<$40$''$ from the quasars and is
significant compared to the galaxy surface density at $\theta$$>$40$''$ in the
same fields.  The other component appears roughly uniform 
to $\theta$$\sim$100$''$ and is significant compared
to the galaxy surface density seen in random-field surveys in the literature.

The $r$$-$$K$ color distributions of the excess galaxy populations are
indistinguishable, and are significantly redder than the color distribution of
the field population.  The excess galaxy population is consistent with being
predominantly early-type galaxies at the quasar redshifts, while there is no
evidence that it is associated with intervening \mg2\ absorption systems.
The average excess within 0.5$h_{75}^{-1}$~Mpc ($\sim$65$''$) of the quasars
corresponds to Abell richness class $\sim$0 compared to the galaxy surface
density at $>$0.5$h_{75}^{-1}$~Mpc from the quasars, and to Abell richness
class $\sim$1.5 compared to that from the literature.  
We estimate --0.65$_{-0.55}^{+0.41}$ magnitudes of evolution in $M_K^*$ to
$\overline{z}$=1.67 by assuming the excess galaxies are at the quasar redshifts.

We discuss the spectral energy distributions (SEDs) of galaxies in fields with
data in several passbands.  
Most candidate quasar-associated galaxies are consistent with 
being 2--3~Gyr old early-types at the quasar redshifts of $z$$\sim$1.5.
However, some objects have SEDs consistent with being 4--5~Gyr old 
at $z$$\sim$1.5, and a number of others are consistent with $\sim$2~Gyr old
but dust-reddened galaxies at the quasar redshifts.
These potentially different galaxy types suggest there may be considerable
dispersion in the properties of early-type cluster galaxies at $z$$\sim$1.5.  
There is also a population of galaxies whose SEDs are best modelled by
background galaxies at $z$$\gtrsim$2.5.

\end{abstract}


\section{Introduction}	\label{intro}

With continuing advances in instrumentation now enabling detailed studies of
very faint and distant galaxies, it is useful to seek efficient methods to find
galaxies and clusters at \z$>$1.  One such possible method is to look for
galaxies associated with quasars, specifically the radio-loud quasars (RLQs)
which comprise $\lesssim$10\% of the quasar population.  

Radio-quiet quasars (RQQs) are rarely found in clusters at any redshift, but
$\sim$35\% of intrinsically luminous (M$_{\rm B}$$<$$-$25) RLQs are located in
clusters of Abell richness class 0--1 (and occasionally 2) at $z$=0.5--0.7
(\cite{yg87}).  
These quasar host clusters typically have anomalously low X-ray luminosities
$L_{\rm X}$ (\cite{hal95}; \cite{hal97}) and velocity dispersions
$\sigma_{\rm v}$ for their richnesses, and thus may be younger and less
virialized than optically-selected clusters (\cite{egy91}; hereafter EGY91).
In addition, RLQ environments are known to evolve rapidly and differentially.  
At $z$$<$0.5 only low luminosity RLQs are seen in richness 1 clusters,
whereas at $z$=0.5--0.7 both high and low luminosity RLQs can be found in such
environments.  A similar effect is seen for FR~II radio galaxies (\cite{hl91}).
EGY91 model this evolution by postulating that galaxy
interactions (more frequent in young, low-$\sigma_{\rm v}$ clusters) are
largely responsible for creating and fueling RLQs, which then fade
as the interaction rate decreases in their evolving, virializing host clusters.
Yee \& Ellingson (1993) suggested that if the cluster formation
rate dropped at $z$$\sim$0.7, most quasars we see in rich clusters at $z$$<$0.7
would be old ones fading on the host cluster's dynamical timescale, and that
luminous RLQs might be found in rich clusters at $z$$>$0.7.
The outstanding feature of quasar evolution is the sharp peak
in space density and/or luminosity at $z$$\sim$2--3 (\cite{sha96}),
but little work has previously been done on quasar environments at z$>$0.7.

Some RLQs show possible additional evidence for being located in 
rich environments, in the form of an excess number of ``associated''
\civ\ (\cite{fol88}) or \mg2\ (\cite{abe94}) absorption systems
within $\pm$5000~km~s$^{-1}$ of the quasar redshift.
There is a tendency for associated absorption to be preferentially found in
steep spectrum sources.  Anderson \etal\ (1987) quote a 2$\sigma$ preference
for associated \civ\ in the unpublished Radio-Loud Survey of Foltz \etal.
This can be seen in Fig.~3 of Foltz \etal\ (1988):  16 of 22 RLQs
with strong associated \civ\ absorption have steep radio spectra, whereas
only $\sim$10 would be expected.  As discussed in Aldcroft, Bechtold \& Elvis
(1994), this dependence is also suggested by the studies of Foltz \etal\ (1986)
and Sargent, Boksenberg \& Steidel (1988).
Associated absorption systems may arise in gas expelled at high velocity from
the quasars or in galaxies in clusters at or near the quasar redshifts.  
In the last few years detailed spectroscopy has shown that some associated
\civ\ systems are almost certainly intrinsic to the quasars,
as discussed in the introduction to Paper 1 (\cite{hgc98}).
However, it remains possible that a substantial fraction of such quasars
reside in clusters which produce associated absorption.  Spectroscopic and
imaging approaches are complementary ways to investigate this question.

\begin{figure} 
\epsscale{0.65}
\plotone{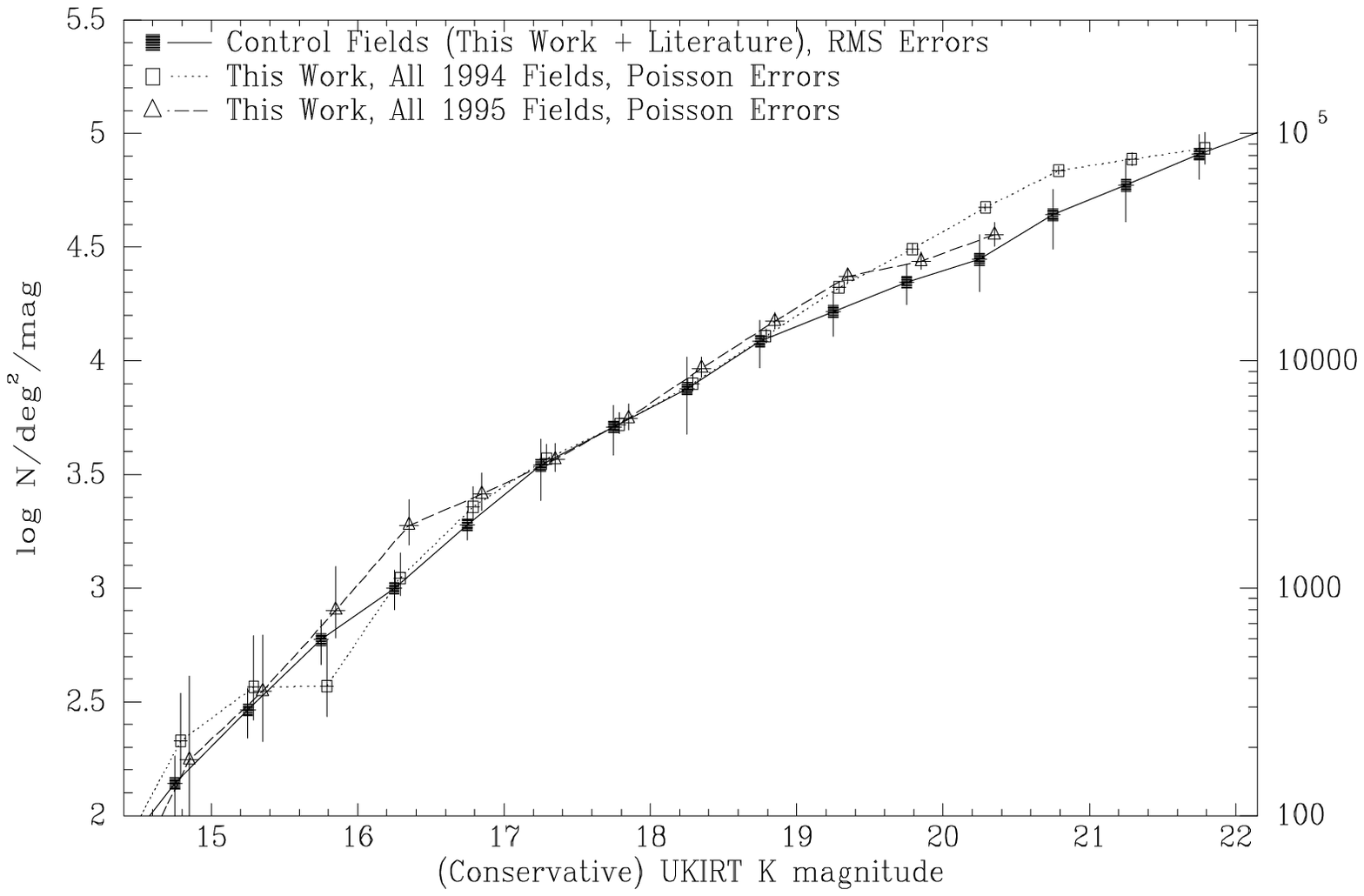}
\caption[Conservative $K_{UKIRT}$ N(m) Relation: Our Data vs. Control Field Ave
age]{
\singlespace
The conservative $K_{UKIRT}$ N(m) relations for our 1994 and 1995 KPNO 4m run
data are plotted as dotted and dashed lines respectively.
The area-weighted average of our conservative-magnitude control fields and all
published random-field imaging surveys (corrected to $K_{UKIRT}$) is plotted
as the solid line.  RMS uncertainties are plotted for the N(m) values and
formal uncertainties for the magnitude bin centers.
}\label{fig_nmsys2vslit}
\end{figure}

In 1994 we embarked upon a project to study the environments of 
RLQs to $z$=2.0, to study any correlations between RLQ environment
and quasar properties such as associated absorption, and to study any examples
of high-redshift galaxies and/or clusters found in RLQ fields.  
Paper 1 (\cite{hgc98}) describes our sample and observations as well as our
data reduction and cataloguing techniques.  
We observed $\lesssim$3$'$$\times$3$'$ fields around 33 RLQs with \z=1--2 and
two control fields with typical 3$\sigma$ limits of $r$=25.5,
\ks=20.5 for $z$$<$1.4, and \ks=21 for $z$$>$1.4.  
An unusual aspect of our data reduction procedure is the creation of images
with constant RMS noise from mosaiced images with exposure times that vary
from pixel to pixel.  A 3$\sigma$ detection thus represents a brighter galaxy
at the edge of an image than in the center, but as long as the exposure time is
everywhere $\geq$36\% of the maximum, the average 5$\sigma$ magnitude limit
will be greater than the 3$\sigma$ detection limit across the entire field.
Object detection was done on RMS-weighted sums of the $r$, \ks, and $J$ data
(where available).  Photometry was done in all filters using the detection
aperture from the summed image.  We use FOCAS total magnitudes and colors
determined from FOCAS isophotal magnitudes.
The useful area surveyed
consists of 221~arcmin$^2$ of RLQ fields and 19~arcmin$^2$ of control fields.
Our control field galaxy counts (and color distribution; 
see Appendix~\ref{litcf})
are consistent with the literature average, but the statistical uncertainties 
are unacceptably large for our purposes, so we also assembled a comprehensive 
set of random-field galaxy surveys from the literature for comparison.
One complication arises from comparison with this literature sample: our 1994
and 1995 photometric solutions produce a $K$-band galaxy number-magnitude
relation N(m) slightly higher than the literature at bright magnitudes.
Given the possible systematic differences between the various $K$-band surveys,
we conduct some of our analyses using a {\em conservative} magnitude scale
(defined so that our number counts exactly match the literature around \k=17)
as well as our original {\em liberal} magnitude scale.
Both scales use UKIRT $K$ magnitudes; the only differences between them
are systematic offsets:  on the conservative scale the 1994 magnitudes
(mostly \z$>$1.4 objects) are 0\fm06 fainter than the liberal scale,
and the 1995 magnitudes (mostly \z$<$1.4) are 0\fm12 fainter, for an average
difference of $\sim$0\fm08 given the numbers of galaxies in each subsample.  
As seen in Figure~\ref{fig_nmsys2vslit}, even under our conservative magnitude
scaling the galaxy counts at \ks$\gtrsim$19 in 31 RLQ fields with good data
are higher than those of field surveys; i.e., 
{\em there is an excess of galaxies in our combined RLQ fields.}

\begin{figure}
\plotone{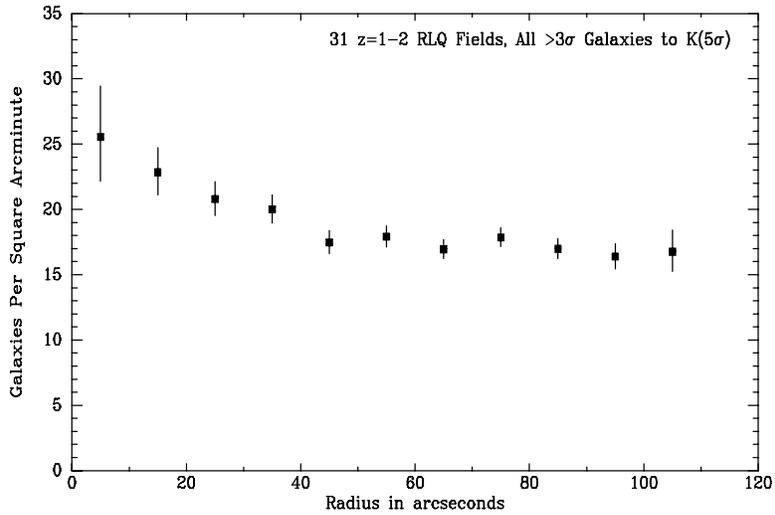}	
\caption[Radial Distribution of Galaxies in All Fields]{
\singlespace
Radial distribution of galaxies relative to the quasars.  All galaxies detected
at $\geq$3$\sigma$ down to the average 5$\sigma$ magnitude limit in each field
are included.  Error bars are calculated for the number of objects in each bin
as per Gehrels (1986).  The large uncertainties are from small number
statistics:  at small radii the bin area is small, and at large radii only part
of the full annulus is imaged, so the number of galaxies detected is small in
both cases.
}\label{fig_rpall1}
\end{figure}

This paper analyzes the excess galaxy population found in Paper 1.
In \S\ref{assoc?} we show that there is an excess of galaxies spatially
concentrated around the quasars themselves (see Figure \ref{fig_rpall1}).
In \S\ref{cmdiag} we show that the excess does not have the same \rks\ color
distribution as the field population described in Appendix~\ref{litcf}
(see Figure \ref{fig_bin1darea_alllibvscfs}).
In \S\ref{richness} we quantify the strength of the clustering under the
assumption it is associated with the quasars and look for correlations
between clustering strength and various quasar properties.
In \S\ref{klf} we estimate the amount of luminosity evolution bright
galaxies have undergone in the $K$-band since \z=1--2.
In \S\ref{seds} we compare the spectral energy distributions of galaxies in
several fields with the predictions of various galaxy spectral evolution models.
We summarize our major results in \S\ref{conclude2}.  
Some additional details are found in Hall (1998).

\section{Are the Excess Galaxies Associated with the Quasars?}  \label{assoc?}

We consider whether the excess galaxies are plausibly physically associated
with the quasars by examining the projected radial distribution of galaxies
around the quasars and the galaxies' distribution in color-magnitude space.
An alternate possibility is that they are intervening galaxies present because 
they trace large-scale matter fluctuations which weakly lens the quasars, 
resulting in quasars from radio catalogs with bright flux limits preferentially
having galaxy excesses around them (``magnification bias"; see \cite{bmm97} 
and references therein).
We observe 134 galaxies in the magnitude range \Kuk=14.5--17 under our
conservative magnitude scaling, and 150 under our liberal magnitude scaling.
The average literature counts compiled in \S 4 of Paper 1
predict 120.3$\pm$22.6 galaxies in our survey area, where the uncertainty is
the RMS scatter among the different surveys in the literature.  Thus the excess
of bright \Kuk$<$17 galaxies in our fields is not statistically significant,
and magnification bias is unlikely to be a significant effect in our sample.

\subsection{Projected Radial Distributions of Galaxies Relative to Quasars} \label{radprof}

If the faint excess galaxies in our fields are physically associated with the
quasars, they are likely to lie preferentially near the quasars on the sky.
To test this we examine the projected radial distribution of galaxies around
the quasars, despite its insensitivity to clusters or groups at the quasar
redshift but not centered on the quasars.  A matched filter (\cite{pos96}) or
cell count (\cite{lp96}) technique could be used to search for off-center
clusters, but would be difficult given the small size of our fields.
However, the galaxy excess over each entire field
should still reflect the presence of off-center clusters.

For this analysis we excluded the fields of Q~0736$-$063
(uncertain stellar contamination) and Q~1508$-$055 (no $r$ data).
We took all galaxies detected at $\geq$3$\sigma$ down to the average 5$\sigma$
$K$ magnitude limit in the 31 other fields, above which essentially no spurious
detections are expected (see \S 3.6.7 of Paper 1).
We binned the galaxies in 10$''$ annuli centered on the quasar and divided by
the area imaged within each annulus.
No correction was made for loss of objects due to crowding or incompleteness
or for stars fainter than our star-galaxy classification limits,
and we do not count the quasar host galaxy.
None of these effects should significantly bias the radial galaxy distribution.
The results (Figure \ref{fig_rpall1}) show a clear excess of galaxies within
40$''$ of the quasars.
The data at $\theta$$<$40$''$ deviates from the level determined from
the $\theta$$>$40$''$ data at the 99.999\% significance level (4.5$\sigma$,
assuming a gaussian probability distribution).

We use the reduced chi-squared $\chi^2_{\nu}$ to quantify the deviation
from a uniform radial distribution.  Here $\nu$ is the number of radial
bins used minus the number of parameters determined from the data.
We use $\chi^2_{\nu}$ both for the deviation from the literature across
the entire field and for the deviation of the $\theta$$<$40$''$ data from the
$\theta$$>$40$''$ data.  Since the $\chi^2_{\nu}$ test requires binning
and does not distinguish over- and under-densities in individual bins,
we also use the Kolmogorov-Smirnov test, which requires no binning but 
is less sensitive to differences in the tails of the
distributions (small or large radii) than in the middle.

To investigate the magnitudes of the galaxies producing the observed excess at
$\theta$$<$40$''$, we take all galaxies with \ks$<$17 and repeat our analysis.
The results 
are consistent with a uniform distribution.
This is a strong indication that at least some of the faint excess galaxies in
these fields are associated with the quasars, since both bright (i.e., \z$<$1)
and faint galaxies should be involved if intervening lensing galaxies 
caused the excess at $\theta$$<$40$''$.

\subsubsection{Dependence on Quasar Redshifts}  \label{rp_qz}

\begin{figure}
\plotone{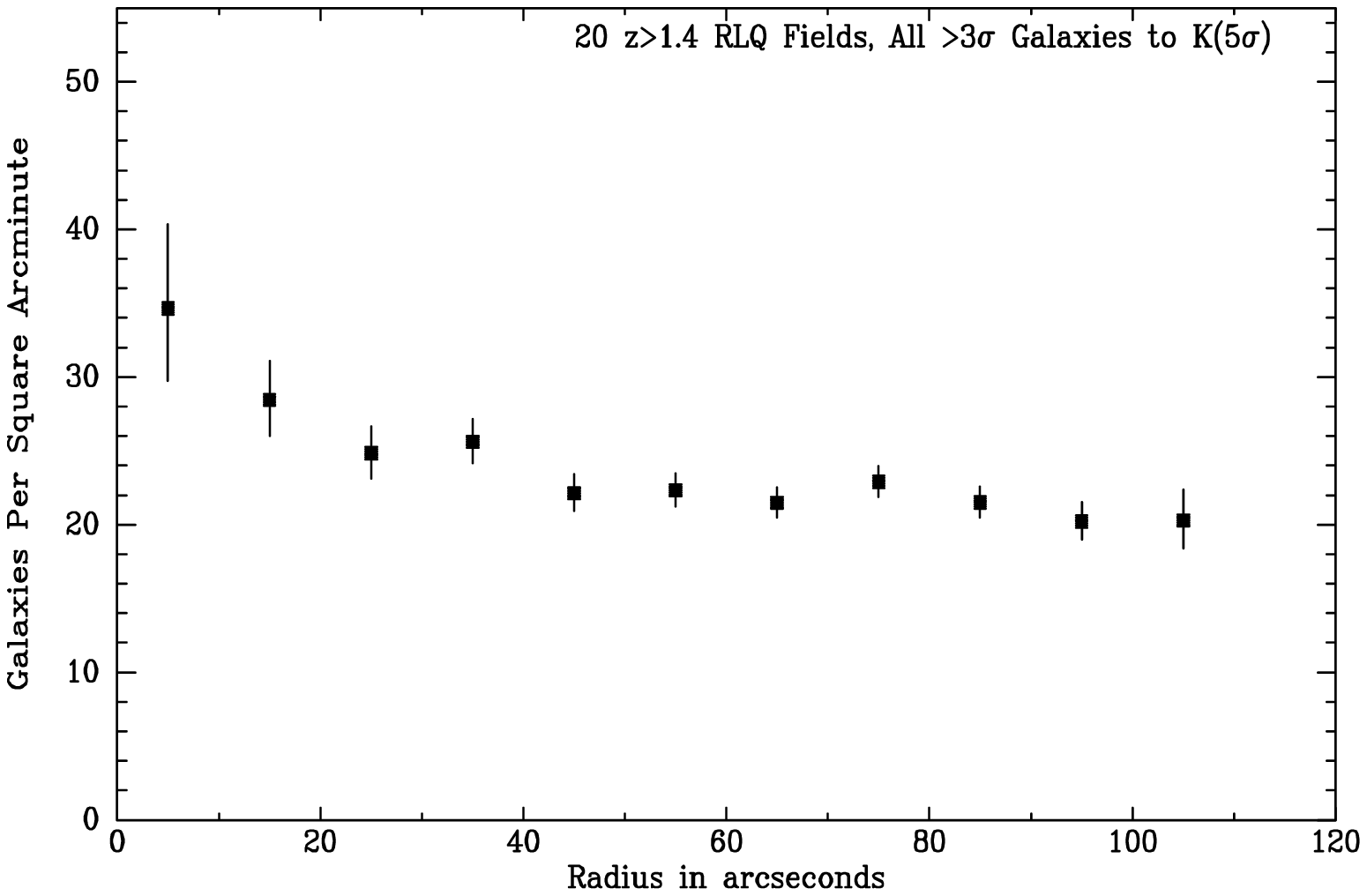}        
\caption[Radial Distribution of Galaxies in $z$$>$1.4 RLQ Fields]{
\singlespace
Radial distribution of galaxies in the $z$$>$1.4 fields.
All galaxies detected at $\geq$3$\sigma$ down to the average
5$\sigma$ magnitude limit in each field are included.
}\label{fig_rphiz_3sigcon}
\end{figure}

\begin{figure}
\plotone{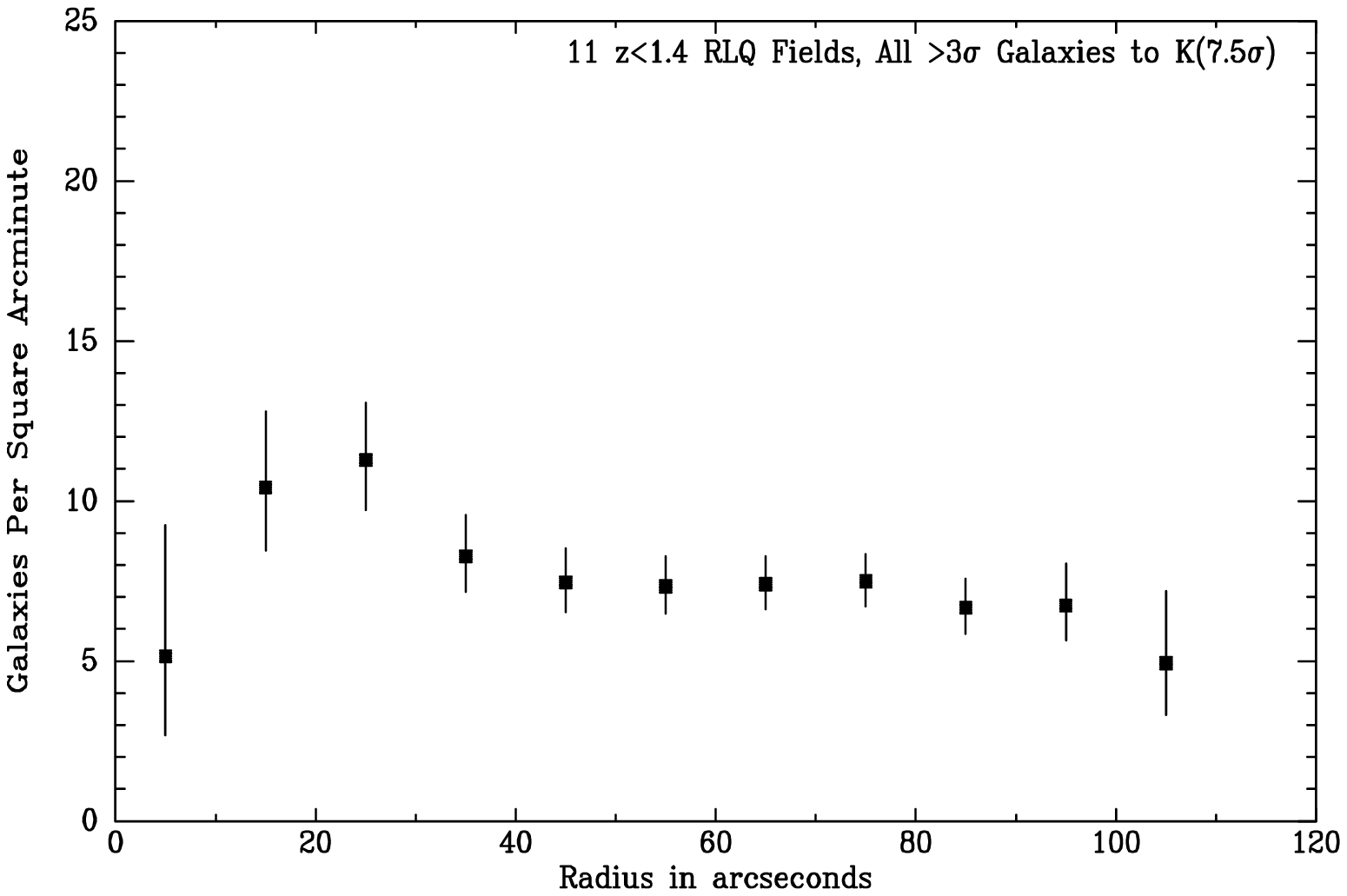}        
\caption[Radial Distribution of Galaxies in $z$$<$1.4 RLQ Fields]{
\singlespace
Radial distribution of galaxies in the $z$$<$1.4 fields.
All galaxies detected at $\geq$3$\sigma$ down to the average
7.5$\sigma$ magnitude limit in each field are included.
}\label{fig_rpmidz_7.5sig}
\end{figure}

To search for any dependence of the galaxy excess on the quasar redshifts, we
split the quasar sample into \z$<$1.4 and \z$>$1.4 subsamples.  
This is a natural division since the two subsamples were observed almost
entirely in different observing runs, and so any systematic magnitude scale
offsets will not affect comparisons within them.
Also, the \z$>$1.4 imaging reaches deeper than the \z$<$1.4 imaging in an
absolute sense as well as relative to the estimated brightest cluster galaxy
magnitude at the average subsample \z.  
Repeating our analysis for low- and high-redshift subsamples with
equal numbers of quasars does not change our results in either a qualititative
or a significant quantitative sense.
Figure \ref{fig_rphiz_3sigcon} shows the 
ensemble radial profile for the high-redshift subsample,
which deviates from the uniform level of the $\theta$$>$40$''$ bins at
99.83\% (3.15$\sigma$) significance (from the K-S test).
The data at $\theta$$<$40$''$ deviates from the $\theta$$>$40$''$ prediction at
the 99.99\% (3.9$\sigma$) level.
The amplitude and spatial profile of the central excess are both
consistent with it being composed of galaxy clusters at the quasar redshifts.
For the $z$$<$1.4 RLQ fields, we plot only objects brighter than 0\fm45 
above the average 5$\sigma$ $K$ limit, equivalent to 7.5$\sigma$ detections,
to ensure uniform detection sensitivity at all radii in spite of different
exposure times.  
Figure \ref{fig_rpmidz_7.5sig} shows a possible overall excess within 30$''$,
but no excess in the innermost 10$''$.  The overall dataset deviates from the 
$\theta$$>$40$''$ prediction at only 65\% significance 
(1$\sigma$) and the $\theta$$<$40$''$ data at only the 97\% (2.2$\sigma$)
level, due to the large uncertainties (our \z$<$1.4 field data is shallower
than our \z$>$1.4 field data).
The $\theta$$<$10$''$ deficit persists even in the complete catalog 
of 3$\sigma$ detections in either $r$ or \ks\ and is not caused by
incompleteness due to seeing or inappropriate dithering.
This deficit may indicate that the $z$$<$1.4 quasars are not centered in
any putative host clusters, but excluding the innermost point does not
raise the significance of the $\theta$$<$40$''$ excess above 3$\sigma$.
If the $\theta$$<$40$''$ excess is real, the $\theta$$<$10$''$
deficit argues against it being due to foreground lensing galaxies.

\subsubsection{Large-Scale Excess}  \label{rp_lgscale}

\begin{figure}
\epsscale{1.0}
\plotone{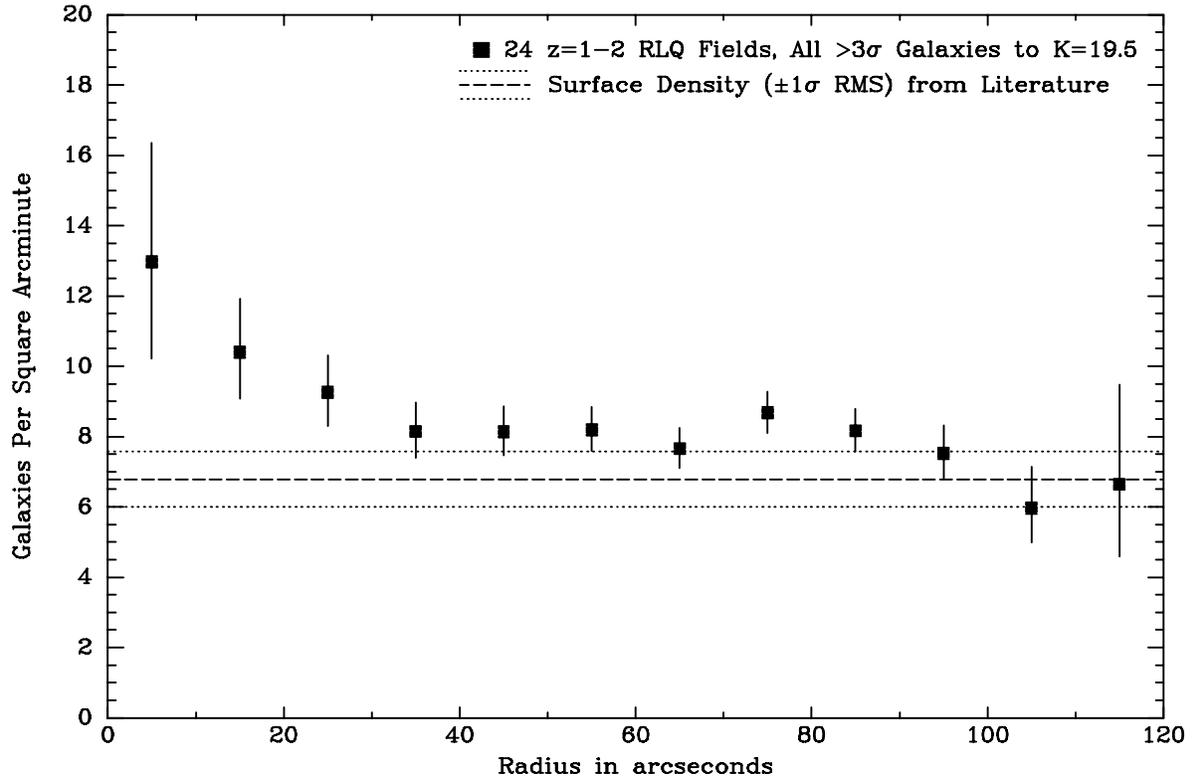}    
\caption[Radial Distribution of Galaxies in All Fields to $K$=19.5]{
\singlespace
Radial distribution of galaxies relative to the quasars in all fields which
reach at least \k=19.5.  All galaxies detected at $\geq$3$\sigma$ down to the
average 5$\sigma$ magnitude limit in each field are included.  
}\label{fig_rpall_k19.5}
\end{figure}

The excess galaxies in the central $\sim$40$''$ radius region
cannot explain the entire excess observed in our fields.  
In Figure \ref{fig_rpall_k19.5} we plot the radial distribution of galaxies to
a fixed limit of \Kuk=19.5 along with the surface density and $\pm$1$\sigma$
RMS dispersion derived from the average published literature counts.
Despite the large uncertainties, there are apparently two components to 
the excess galaxies:  the central component ($\theta$$<$40$''$) and a
large-scale component extending to $\theta$$\sim$100$''$.
(At \z$\sim$1.5 in our \ho=75, \qo=0.1 cosmology, 40$''$ is $\sim$0.3\hn1~Mpc
and 100$''$ is $\sim$0.75\hn1~Mpc.)
Could this large-scale component be spurious?
The RMS uncertainties on the combined literature counts are in good agreement
with galaxy clustering predictions based on single surveys (\cite{cea97};
\cite{reh97}), so the range shown for the literature should be accurate.
We do not expect many spurious objects above the 5$\sigma$ limits in our fields.
We have neglected contamination by stars fainter than our star-galaxy
classification limits, but also incompleteness corrections
to our fields (though not to the literature data).
If we include both effects, the galaxy surface density is essentially
unchanged to \Kuk=20 and actually increases (by $\lesssim$3\%) to \Kuk=20.5.

A simple check of the need for a large-scale excess can be made:
from Figure \ref{fig_rpall1}, the central excess at $\theta$$<$40$''$
above the surface density defined by the $\theta$$>$40$''$ points is only 
$\sim$10\% of the total counts, or 0.04~dex, but Figure~\ref{fig_nmsys2vslit}
shows that our excess counts are $\sim$0.1~dex (25\%)
above the literature average at 19$<$\Kuk$<$21.5.  Thus a large-scale excess
of amplitude roughly equal to the central excess must exist in our fields.
Supporting evidence for the reality of the large-scale excess can be found in
its \rks\ color distribution (Figure~\ref{fig_bin1dnorm_alllt40libvsgt40}),
which is indistinguishable from that of the excess at \th$<$40\arcsec.

\subsubsection{Ubiquity of the Excess Galaxies}  \label{rp_ubiquity}

\begin{figure}
\epsscale{0.60}
\plotone{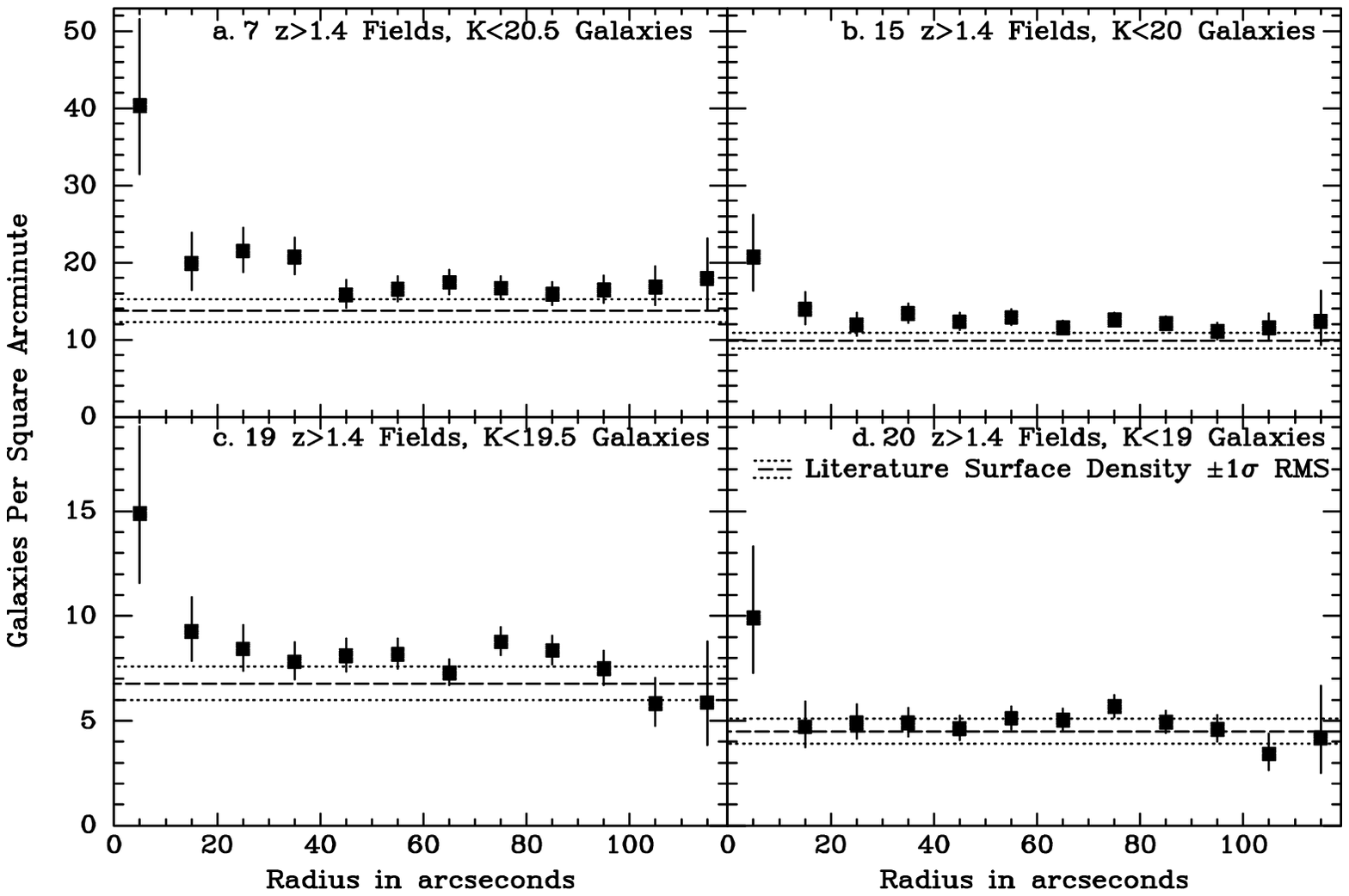}	
\caption[Radial Distribution of Galaxies in Different Magnitude Ranges in $z$$>$1.4 RLQ Fields]{
\singlespace
Radial distribution of galaxies in the $z$$>$1.4 RLQ fields.
Dashed line in both figures is the surface density from the average published
literature counts, and dotted lines are the $\pm$1$\sigma$ RMS uncertainties.
a.  All galaxies detected at $\geq$3$\sigma$ down to \Kuk=20.5 in the 7 
$z$$>$1.4 RLQ fields with 5$\sigma$ limits that deep.
b.  All galaxies detected at $\geq$3$\sigma$ down to \Kuk=20 in the 15 
$z$$>$1.4 RLQ fields with 5$\sigma$ limits that deep.
c.  All galaxies detected at $\geq$3$\sigma$ down to \Kuk=19.5 in the 19 
$z$$>$1.4 RLQ fields with 5$\sigma$ limits that deep.  d.  All galaxies
detected at $\geq$3$\sigma$ down to \Kuk=19 in all 20 $z$$>$1.4 RLQ fields.
}\label{fig_rphiz_new}
\end{figure}

\begin{figure}
\plotone{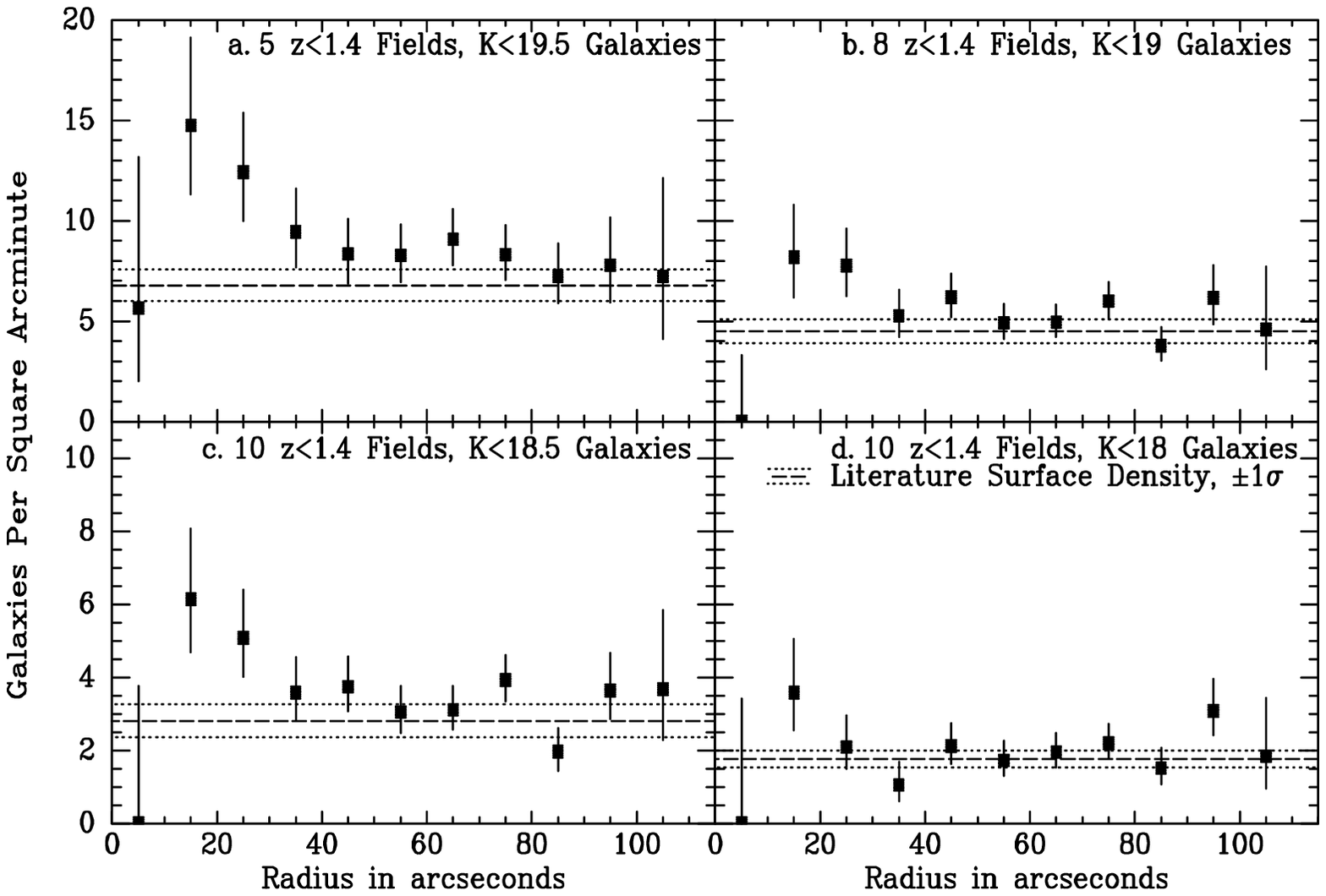}    
\caption[Radial Distribution of Galaxies in Different Magnitude Ranges in $z$$<$1.4 RLQ Fields]{
\singlespace
Radial distribution of galaxies in the $z$$<$1.4 RLQ fields.
Dashed line in both figures is the surface density from the average published
literature counts, and dotted lines are the $\pm$1$\sigma$ RMS uncertainties.
a.  All galaxies detected at $\geq$3$\sigma$ down to \Kuk=19.5 in the 5
fields with 5$\sigma$ limits that deep.
b.  All galaxies detected at $\geq$3$\sigma$ down to \Kuk=19 in the 8
fields with 5$\sigma$ limits that deep.
c.  All galaxies detected at $\geq$3$\sigma$ down to \Kuk=18.5 in the 10
$z$$<$1.4 RLQ Fields fields with 5$\sigma$ limits that deep.
d.  All galaxies detected at $\geq$3$\sigma$ down to \Kuk=18 in all 11 fields.
}\label{fig_rpmidz_new}
\end{figure}

We consider various subsamples of our data to see how common the 
excess galaxies are.
We plot the radial distribution of galaxies in our \z$>$1.4 fields to various
limits in Figure \ref{fig_rphiz_new}.
Using the
average surface density from the $\theta$$>$40$''$ data, the $\theta$$<$40$''$
excess has significance 99.7\% (3$\sigma$) at \K$<$20.5,  
and $\leq$85\% ($\leq$1$\sigma$) at \K$<$20 or brighter.
Using the average literature surface density and RMS, the overall excess 
(central and large-scale) has significance 99.9\% (3.3$\sigma$) at \K$<$20.5
or \K$<$20 and $<$95\% (2$\sigma$) at \K$<$19.5 or \K$<$19.
Using our conservative magnitude scale effectively moves the literature 
surface density upward slightly.  The overall excess then has
significance of only 96--97\% (2--2.2$\sigma$) at \K$<$20.5 or \K$<$20.
Thus for our \z$>$1.4 fields alone,
the central excess is only significant if the magnitude limit reaches \K$=$20.5,
while the large-scale excess is only significant if the magnitude limit is
\K$=$20 or fainter, and then only under our liberal magnitude scale.
Due to shallower magnitude limits and fewer fields, the uncertainties are
larger at \z$<$1.4 (Figure \ref{fig_rpmidz_new}).
The overall excess compared to the average
surface density from the literature is only significant at the 95\% (2$\sigma$)
level at best, and even less under our conservative magnitude scaling.
We have confidence in the reality of this large-scale excess despite
these low formal significances, since the
limiting factor in determining the significance 
is the intrinsically large field-to-field RMS of $K$-band galaxy counts.

The central excess may be 
produced by only a few of the fields, but the large-scale excess is not.
If we exclude the 5 (out of 20) \z$>$1.4 fields with highest $\chi^2$,
the central excess drops below 3$\sigma$\ significance.  Thus it may be
produced by as few as $\sim$25\% of the fields (a $very$ uncertain fraction).
However, if we take the 15 \z$>$1.4 fields which reach \K=20 and remove 
the 5 fields with highest $\chi^2$, the resulting overall excess is still 
significant at the 99.7\% (3$\sigma$) level, as measured by both the $\chi^2$
and K-S tests.  Thus the large-scale galaxy excess is not produced solely by
the same fields which may contribute most of the central galaxy excess.
At least $\sim$50\% of our RLQ fields contribute to the large-scale
galaxy excess.  Both fractions may be higher since removing fields
will reduce the S/N and significance (to $<$3$\sigma$) even
when all fields contribute to the large-scale excess.

Only three of our fields might have a large-scale excess produced by very
low redshift galaxy associations listed in NED\footnote{The NASA/IPAC
Extragalactic Database (NED) is operated by the Jet Propulsion Laboratory,
California Institute of Technology, under contract to NASA.} within 18$'$:
Q~1018+348, 7\farcm4 from Abell 982 (no published redshift0; Q~1221+113,
1\fdg7 from the center of the Virgo cluster and 7\farcm4 from NGC~4352,
and Q~2230+114, 5\farcm5 from NGC~7305.  Excluding them does not
significantly reduce the $\chi^2_{\nu}$ of any fit.
Another remote possibility is that faint galaxies associated with the
supergalactic plane (\cite{dev75}) might affect our counts.
Using NED, we found the average absolute supergalactic latitudes to be
31$\pm$17$\arcdeg$ for our \z=1--1.4 RLQ subsample,
27$\pm$16$\arcdeg$ for our \z=1.4--2 RLQ subsample, and
23$\pm$14$\arcdeg$ for published random-field $K$-band surveys.
Thus even if faint galaxy counts do correlate
with supergalactic latitude, there should be no systematic offset between our 
counts and those of the published literature data due to such an effect.

\subsection{Correlations with Quasar Properties}      \label{radprof_correlate}

Finally, we examine the dependence of the central excess on various quasar
properties using the $\chi^2$ test.  Ordering our sample on absolute $V$
magnitude and splitting it in half, 
we find a 4$\sigma$ 
$\theta$$<$40$''$ excess for the more luminous RLQs ($M_V$=--27.21$\pm$0.61)
but only a 2$\sigma$ excess for the fainter RLQs ($M_V$=--25.31$\pm$0.77).
However, these subsamples have average redshifts $\overline{z}$=1.64$\pm$0.27
and 1.32$\pm$0.28 respectively.  The central excess is more significant
in our \z$>$1.4 subsample 
than in our \z$<$1.4 subsample.  
Thus we cannot say whether the primary dependence is on $M_V$ or redshift.
In fact, the apparent redshift dependence may be due to the fact that our
\z$>$1.4 data reaches an average of 2\fm4 below $K_{BCG}$ ($K_{BCG}$ is the
estimated brightest cluster galaxy magnitude at the quasar redshifts), as
opposed to 2\fm1 below $K_{BCG}$ for our \z$<$1.4 data (see \S\ref{calcn05}).
This is in turn because our 1995 IRIM data (mostly \z$<$1.4 objects) has higher
backgrounds and poorer seeing than our 1994 IRIM data (mostly \z$>$1.4 objects).

Similarly splitting the sample in half at radio power log~$P_{rad}$=27.5~W/Hz,
we find a $>$4$\sigma$ $\theta$$<$40$''$ excess for more powerful RLQs
(log~$P_{rad}$=27.79$\pm$0.21) but only 2$\sigma$ for less
powerful ones (log~$P_{rad}$=27.14$\pm$0.37).
The average redshifts of these subsamples are 1.46$\pm$0.32 and 1.49$\pm$0.32
respectively, and the average $M_V$ are --26.20$\pm$1.40 and --26.26$\pm$0.95,
so this does seem to be a dependence on radio power.

If we compare the 11 flat-radio-spectrum and 19 steep-radio-spectrum objects,
we find a 3.6$\sigma$ $\theta$$<$40$''$ excess around the steep-spectrum
objects but only a 1.65$\sigma$ excess around the flat-spectrum ones.  
These two subsamples have very well-matched redshift, \mabs, and 
\prad\ distributions, so again the dependence seems to be on radio spectrum.  
This is supported by the results of
splitting the sample between sources with and without strong radio lobes
(types FRII or T and types C or CE, respectively, in Table 1 of Paper 1),
subsamples which are 84\% and 25\% steep-spectrum objects
respectively.  For strong-lobed sources we see a 3.4$\sigma$ $\theta$$<$40$''$
excess, while for weak-lobed sources the excess is only 2.6$\sigma$.

Splitting our sample according to the quasars' associated absorption properties,
we find a 2.8$\sigma$ $\theta$$<$40$''$ excess in 13 objects with associated
absorption and a 2.4$\sigma$ excess in 9 objects with no associated absorption.
These subsamples are slightly mismatched in $M_V$,
but well matched in \z\ and \prad.
Detecting excesses around steep-spectrum RLQs but not ones with associated
absorption may be surprising in light of the $\gtrsim$2$\sigma$ tendency for
steep-spectrum RLQs to preferentially show associated absorption (see the 
introduction to Paper 1).  However, this tendency is not strong enough and our
subsamples are not large enough to show a significant contradiction.

\subsection{Summary and Discussion:  Radial Profiles}    \label{radprof_summary}

We see an excess of galaxies at $\theta$$<$40$''$ from these 31
RLQs compared to the $\theta$$>$40$''$ background level.  
This central excess is significant at the $\sim$99.995\% ($\sim$4$\sigma$)
level and consists of $K$$>$17 galaxies.
It is seen at 3--4$\sigma$ significance in the \z$>$1.4 fields,
but only at $\sim$2$\sigma$ significance in the \z$<$1.4 fields which are
shallower and fewer in number.
The central excess at \z$>$1.4 may be produced by as few as 5 of the 20
fields, but it is not due to one or two extreme outliers.
There is an additional large-scale galaxy excess extending to
$\theta$$\sim$100$''$ detectable in our \z$>$1.4 fields with 3.3$\sigma$ 
significance at $K$$<$20 or $K$$<$20.5.  
It may also be present in our \z$<$1.4 fields, but the larger uncertainties
there preclude a detection above 2$\sigma$.
We consider various possible errors and conclude that the large-scale excess is
real at the given significance levels, and that it is not produced solely by
the same fields which may contribute most of the central galaxy excess.
The $\theta$$<$40$''$ excess does not depend on the presence of associated
absorption, but seems to be stronger (4$\sigma$ vs. 2$\sigma$) for the more 
radio-powerful RLQs, and may be stronger (3.6$\sigma$ vs. 1.65$\sigma$) for
steep-radio-spectrum ($\alpha_r$$\geq$0.5) RLQs than for flat-spectrum ones.

Could the excess galaxies be associated with intervening \mg2\ absorption-line
systems?  This is difficult to quantify this using radial profiles due to the
variation in limiting magnitude between fields and the low formal significances
of the excesses at all but the faintest magnitudes.
However, we show in \S\ref{n05_summary} that the 8 \z$>$1.4 RLQs with known
intervening absorption and the 8 \z$>$1.4 RLQs without it show no differences
in their inferred richnesses.  Thus there is no evidence that galaxies
associated with intervening \mg2\ systems contribute significantly to the
observed galaxy excess.

The large-scale (to $\theta$$\sim$100$''$) excess can plausibly be located
at the quasar redshifts.
At \z=1.4--2.0, a 200$''$ diameter circle corresponds to $\sim$1.5\hn1~Mpc
diameter, which is not an implausibly large scale, especially if
it is connected with the $\theta$$<$40$''$ (0.6\hn1~Mpc diameter) excess which
might be e.g. a cluster core or group embedded in a larger overdensity.
From the PLE model redshift distribution of Roche, Eales \& Hippelein (1997),
we estimate that $\sim$1.5$\pm$0.5\% of field galaxies to \K=20 lie within
$\delta$\z=0.05 of \z=1.70.  
This redshift bin size is large but similar to those in which overdensities have
been spectroscopically confirmed at high \z\ (\cite{dic96b}; \cite{ste98a}).
Our observed galaxy surface density to \K=20 is 1.23$\pm$0.14 times the
literature (Figure \ref{fig_rphiz_new}).  
Thus if all this excess was at the quasar redshifts, it would consitute a
typical galaxy number overdensity of $\sim$14$\pm$10 compared to the
model redshift distribution.
This overdensity is again plausible:  virialized clusters are overdense by 
similar factors on similar scales, the RMS galaxy fluctuation 
on 8\hm1~Mpc scales is $\sigma_8$$\simeq$1 (\cite{lin96b}), and superclusters
have overdensities $\sim$5--40 on $\sim$30\hm1~Mpc scales (\cite{sea98}).
The candidate large-scale structures in these fields are not unprecendented;
they would be similar to the galaxy
overdensities of $\sim$10 spectroscopically confirmed by Deltorn \etal\ (1996;
1997) and Dickinson (1996b) at and/or near the redshifts of three \z$\sim$1
radio galaxies on very similar spatial scales.  
In particular, the overdensity near 3C~324 (\cite{dic96b})
is composed of two clumps or sheets of galaxies separated by 
$\sim$7500~km~s$^{-1}$ in their rest frame, evidence for the existence
of large-scale structures around at least some \z$>$1 radio-loud AGN.
Similarly, the 8$'$-separation \z=0.6 quasar pair 3C~345 and Q~1641+3998
may be embedded in a large-scale galaxy structure (\cite{ey94}).

\section{Color-Magnitude Diagrams}      \label{cmdiag}

Color-magnitude diagrams (hereafter CMDs and denoted as magnitude/color, e.g.
\ks/\rks) can provide useful information on the excess galaxies in our fields.
Since field galaxies contaminate CMDs at all magnitudes and colors, comparison
of quasar-field and control-field CMDs must be made.  
We discuss the literature control field datasets we use in Appendix~\ref{litcf}.

To reduce the uncertainties on the colors, we calculate them using FOCAS
isophotal magnitudes.  We use the same aperture in all filters, namely the
isophotal aperture from the coadded $r$+$K_s$ or $r$+$J$+$K_s$ detection image
for the field.  This also allows for more accurate color limits, since
measuring colors through smaller apertures
allows very faint objects to be detections instead of 3$\sigma$ upper limits.
Comparison of isophotal- and total-aperture colors showed no systematics
and a scatter consistent with photometric uncertainties.

\begin{figure}
\epsscale{0.65}
\plotone{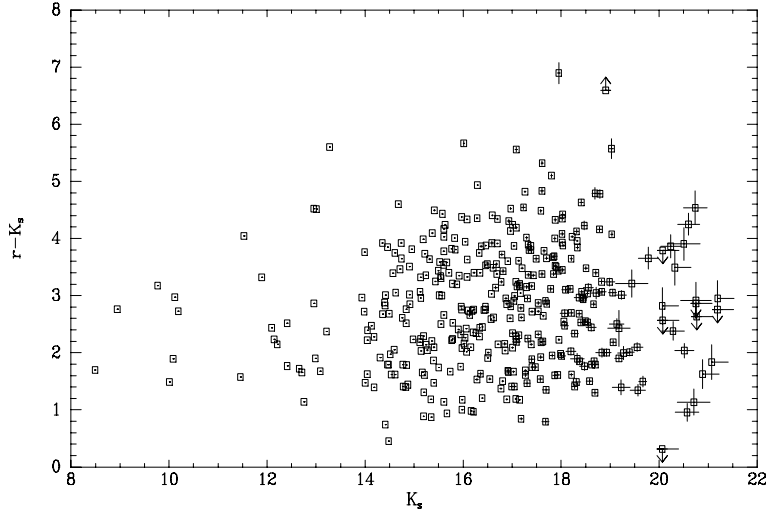} 		
\caption[\ks/\rks\ Color-Magnitude Diagram: Stars]{
\singlespace
Morphologically identified stars' isophotal \rks\ colors vs. total $K_s$
magnitudes.  Error bars are 1$\sigma$ photon noise only.
Faint objects with only rightward error bars on their \ks\ magnitudes
have total magnitude detections of $<$3$\sigma$,
but may have isophotal magnitude detections of $>$3$\sigma$.
All RLQ fields are plotted except Q~1508$-$055 (no $r$ data) and Q~2230+114 
(nonphotometric $R_C$ data).  The 31 remaining fields cover 219.2~arcmin$^2$
and contain 380 stars detected at $\geq$3$\sigma$ in $r$ or $K_s$.
}\label{fig_krk_allstars}
\end{figure}

\begin{figure}
\plotone{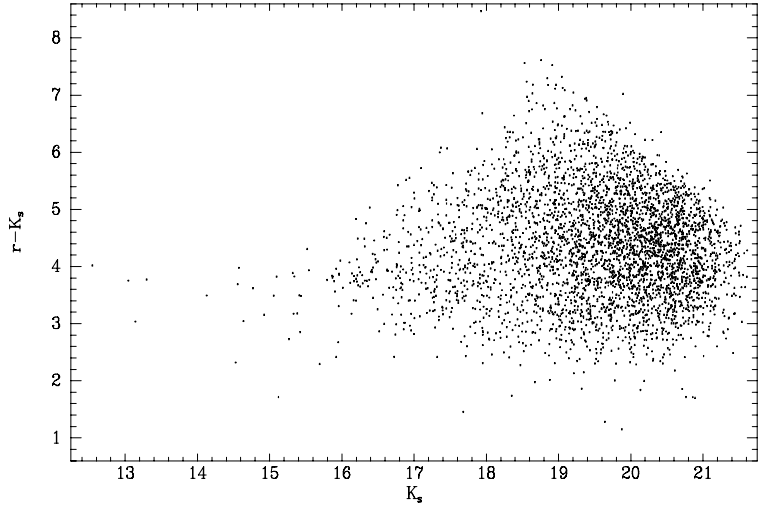} 		
\caption[\ks/\rks\ Color-Magnitude Diagram: Galaxies]{
\singlespace
Galaxy \rks\ colors vs. $K_s$ magnitude.  All RLQ fields are plotted except 
Q~1508$-$055 (no $r$ data), Q~2230+114 (nonphotometric $R_C$ data), 
and Q~0736$-$063 (uncertain stellar contamination).
The 30 remaining fields cover 213.7~arcmin$^2$ and contain 3886 
galaxies detected at $\geq$3$\sigma$ in $r$ or $K_s$ and with $K_s$ brighter 
than the average 5$\sigma$ limits, all of which are plotted here.  
Note the reddest object at $K_s$$\sim$18, $r$$-$$K_s$$\gtrsim$8.4
(3$\sigma$ lower limit).
}\label{fig_krk_allgals}
\end{figure}

Figure \ref{fig_krk_allstars} shows the \ks/\rks\ CMD for stars in 31 RLQ
fields.  Typically the robust star-galaxy separation limit is 
$K_{classlim}$=17.5--18, hence the dropoff in star counts fainter than that.
Stars at $K_s$$>$20 were classified from snapshot $HST$ images.
Figure \ref{fig_krk_allgals} shows the \ks/\rks\ CMD for galaxies in 30 RLQ
fields, with upper and lower limits and error bars omitted for clarity.
A galaxy with a flat spectrum in $f_{\nu}$ has \rks=2.06 with our adopted 
zeropoints (see Appendix A of Paper 1 and Djorgovski \etal\ 1995).

We have statistically corrected the number-magnitude relation at 
\ks$>$$K_{classlim}$ for unidentified faint stars, but how do we statistically
account for them in comparing quasar-field and control-field CMDs?
Stars have a bluer mean \rks\ color than galaxies, and thus could bias
comparisons between fields at different Galactic latitude (and longitude).
We could assume that stars uniformly populate the color range \rks=1--4.5.
However, our $K_{classlim}$ values are faint enough that stellar 
contamination is only a $\sim$5\% effect at fainter magnitudes.
We thus make no correction for stellar contamination in our consideration of
CMDs, but will discuss the effects of it where relevant.

\subsection{One-Dimensional Histograms}	\label{cc1d}
  
One simple method of comparing CMDs is to split them into different magnitude
bins and compare the color histograms in each bin.  In the following discussion
we exclude the fields of Q~1508$-$055 (no $r$ data), Q~2230+114 (nonphotometric
$R_C$ data), and Q~0736$-$063 (uncertain stellar contamination).
At the 0\fm5 binning size we use, the $\sim$0\fm08 difference between our
conservative and liberal magnitude scales is not significant, so we adopt the
latter.  
In Figure \ref{fig_bin1darea_alllibvscfs} we plot the surface density of
galaxies, binned by \rks\ color, from all 30 good RLQ fields (solid lines)
and in the combined opt-IR control fields (dotted lines) discussed in
Appendix~\ref{litcf}.  Smaller histograms represent those galaxies with lower
or upper limits to their colors.
At \Kuk$\gtrsim$18, there is a clear excess of red galaxies with
\rK$>$5 in the quasar fields.
This color and magnitude range is where we would expect to see
bright cluster ellipticals associated with the quasars.

To quantify the significance of any difference between the color distributions
regardless of surface density, taking into account upper and lower limits,
we normalize all histograms separately to unit sum
and use the Peto \& Prentice Generalized Wilcoxon Test as implemented in the
IRAF\footnote{The Image Reduction and Analysis Facility (IRAF) is distributed
by National Optical Astronomy Observatories, operated by the Association of
Universities for Research in Astronomy, Inc., under contract to the NSF.}/STSDAS
task {\sc statistics.twosampt}.
The two \rk\ distributions are different at only 67\% (1$\sigma$) significance
at \Kuk=17--18, but at 99.98\% (3.7$\sigma$) significance at \Kuk=18--19 and at
$>$99.995\% ($>$4$\sigma$) significance in the faintest two bins shown, 
and at \Kuk$>$21 (not shown) for the handful of fields that reach that deep.
Thus the excess galaxy population in these RLQ fields has a redder \rk\ color
distribution than the field population at $>$99.995\% ($>$4$\sigma$) 
significance.
Random field to field variations in \rk\ color distributions are possible
due to variations in galaxy population with environment, but should average
out over many fields.  Thus the excess population of predominantly
red galaxies must somehow be connected with the presence of RLQs in our fields.

The red galaxy excess persists at \Kuk=20--21, in which magnitude range
there is a significant deficit of blue galaxies.  This apparent deficit is
difficult to understand, since the \Kuk$\sim$20.5 faint blue galaxy population
is mostly at \z$<$1 and should be uncorrelated with the presence of \z$>$1
quasars in these fields.  If foreground galaxy structures cause
magnification bias in these fields, one might expect slightly fewer blue
(late-type) galaxies in the fields due to the tendency of red (early-type)
galaxies to preferentially inhabit denser environments, but this tendency
is too weak to explain a complete absence of blue galaxies or why it occurs
only in the faintest magnitude bin.  In addition, the Butcher-Oemler effect
(\cite{bo84}) is seen in clusters to \z$\sim$1 (\cite{rs95}; \cite{lub96}), 
so it is natural to expect that these candidate \z=1--2 clusters should also
have a large ($\sim$40\%) fraction of blue galaxies.

However, this apparent deficit is also difficult to explain away as spurious.  
There is adequate control field data ($\sim$200 galaxies) at \Kuk=20--21 from 
our control fields and the Hubble Deep Field IRIM (HDF-IRIM), Moustakas, and
Djorgovski datasets.  
It is possible that conversion of the HDF-IRIM and Moustakas $V$ and $I$
magnitudes to $r$ causes part of the offset between quasar- and control-field
histograms, but the full $\sim$0\fm5 offset is inconsistent with the scatter
in the conversion (Eq.~\ref{eq_r_from_VI}) and the good agreement between
control-field datasets obtained with different optical filters
(Figure~\ref{fig_bin1dnorm_cfrkvslit}).
More subtle errors involving different instruments and different magnitude
measurement techniques cannot be ruled out, but the two largest control field
datasets in this magnitude range were both obtained on the KPNO 4m with IRIM
and catalogued using FOCAS total magnitudes and isophotal colors.
Using our conservative magnitude scale reduces the deficit only slightly.
The apparent deficit is of objects with $r$=22--24 and \Kuk=20--21,
which is well above the detection limits for the $r$ data 
but within $\sim$0\fm5--1\fm0 of our 3$\sigma$ $K_s$ limits.  
The Eddington bias (\cite{edd13}), the systematic overestimate of faint object
fluxes due to the increase in number counts with decreasing flux, should only
bias our $K_s$ magnitudes brightwards by $\sim$0\fm1--0\fm2 at our 5$\sigma$
$K_s$ limits, using Eq.~7 of Hogg \& Turner (1997).  
However, this error will occur in our control fields as well, and they still
show an excess of blue galaxies compared to our quasar fields.
Any further systematic errors in our magnitude measurements are likely to exist
only at our faintest limits (i.e. affecting only the $K_s$ magnitudes of any
``missing'' objects) and to $underestimate$ the magnitudes of faint objects
(e.g. if FOCAS total magnitudes underestimate the flux for the lowest S/N 
objects compared to brighter objects).  However, any such errors in \ks\ would 
produce a bias toward bluer \rks\ colors, the opposite of what is observed, and
would also occur in our control fields as well as quasar fields.
Lastly, there is no significant deficit of galaxies blue in \jk\ at \k=19--20
or \k=20--21 in our data compared to the HDF-IRIM control field data.
This suggests that the error occurs either in $r$ or in both $J$ and \ks\ if it
is a straightforward magnitude measurement error, but we suspect it is instead
due to a combination of errors.  In any case, the excess of red galaxies
at \Kuk=20--21 is still significant even if we arbitrarily shift the quasar-
and control-field \rK\ color histograms so that they match at the blue end.

In Figures \ref{fig_bin1darea_midzlibvscfs} and \ref{fig_bin1darea_hizlibvscfs}
we plot the surface density of galaxies in each color bin for the \z$<$1.4 and
\z$>$1.4 subsamples separately.
The \z$<$1.4 subsample is noisier, as usual, and is only useful down to \Kuk=20.
The red galaxy excess is definitely present at \Kuk=19--20, and perhaps at
\Kuk=18--19 as well.
The \z$<$1.4 and control field distributions are different at the 99.71\%
(3$\sigma$) significance level at \Kuk=18--19 and at the $>$99.995\%
($>$4$\sigma$) level at \Kuk=19--20.
The \z$>$1.4 and control field distributions are different at the 99.99\%
(3.9$\sigma$) significance level at \Kuk=18--19, and at the $>$99.995\%
($>$4$\sigma$) level at \Kuk=19--20 and \Kuk=20--21.

In Figure \ref{fig_bin1dnorm_alllt40libvsgt40} we plot the fractional \rk\ color
distributions for galaxies at $\theta$$<$40$''$ (solid line) and
$\theta$$>$40$''$ (dotted line) from the quasars.  There is no significant
difference between the histograms.  
The similarity between the color distribution of the excesses at small and
large radii suggests that the large-scale galaxy excess is the same population
as the $\theta$$<$40$''$ excess.

As a check on the reality of the redness of the excess population, we can plot
the \Kuk/\jKuk\ CMD for the five fields with $J$ data.  Figure \ref{fig_kjk_cfs}
shows the \K/\jk\ control field CMD (see Appendix~\ref{litcf}),
and Figure \ref{fig_kjk} shows the \K/\jk\ CMD of the five \z$>$1.4 RLQ fields.
Comparing the two, there appears to be an excess of galaxies in 
the RLQ fields with \k$\gtrsim$19 and \jk$\gtrsim$2.
In Figure \ref{fig_bin1dnorm_kjkvscfs} we plot the normalized \jk\ color 
histograms in four magnitude bins.
The \jk\ distributions are not significantly different at \k$<$19.
They are different at the 98.28\% confidence level (2.4$\sigma$) at \k=19--20 
and at the $>$99.995\% ($>$4$\sigma$ level) at \k=20--21, in both cases due to
the tail at \jk$\gtrsim$2 in the RLQ fields.
Only the HDF-IRIM dataset has $J$ data for objects at \k$>$20.
(Note added in proof:  the recent data of Bershady, Lowenthal \& Koo (1998)
has a \jk\ distribution at \ks=20--21 consistent with HDF-IRIM.)
Thus there is some evidence that the excess galaxy population is redder than
the field population in \jk\ as well as \rk.  
This more subtle difference is not surprising since even to \k=20 most galaxies
are at \z$\leq$2 (\cite{cow96}) and the red envelope of \jk\ colors for
early-type galaxies increases only slowly out to that redshift.

\subsection{Summary:  Color-Magnitude Diagrams}		\label{cmdiag_summary}

The \rk\ color distribution of the faint excess galaxy population is
significantly redder than that of the field population.  
The \jk\ color distribution for the five quasar fields with $J$ data also
shows a red tail not present in the field.  
There is no significant difference between the color distributions of the
$\theta$$<$40$''$ and large-scale excess components.
The colors and magnitudes of both excess populations are thus consistent with
a population of predominantly early-type galaxies at the quasar redshifts.
There is an apparent deficit of blue galaxies in the faintest magnitude bins
which is difficult to understand as either a real effect or a single systematic
error.  However, the red galaxy excess is still significant if we arbitrarily
adjust the \rk\ color histograms so as to eliminate the apparent deficit.

\section{Estimates of Cluster Richnesses}      \label{richness}

We have seen that there is a significant faint galaxy excess in our RLQ fields
with a color distribution redder than that of the field population.  
While only spectroscopy can determine the redshift distribution of the 
galaxy excess for certain, its magnitude, color, and 
spatial distributions are consistent with it being composed of clusters or 
other large structures at the quasar redshifts.  We now assume that 
this is the case 
in order to quantify the richnesses such clusters would have.

Since we have no information on the dynamical state of these structures, our
use of the term ``clusters'' can be somewhat misleading.
Spectroscopy is needed to confirm or deny the hypothesis that these structures
are virialized or virializing.
Even then, the evolution of individual galaxies and clusters 
from \z$>$1 to \z$\sim$0 might make accurate comparison with low-redshift
clusters difficult without detailed comparison with simulations to identify
similar populations of objects at each \z.
We discuss these and similar issues further in \S\ref{n05_summary},
but for now we assume the strength of the galaxy excess in our fields can be
compared more or less directly with that of low-redshift clusters.

Abell (1958) defined a cluster's richness as the number of member galaxies 
above background no fainter than 2\fm0 below the third brightest
cluster galaxy and within 1.5$h^{-1}$~Mpc of the center center.
This criterion is difficult to apply directly at high \z\ due to the
large and uncertain background correction required over 3$h^{-1}$~Mpc diameter
and also to the uncertainty in identifying the third brightest cluster galaxy.
Thus we use the alternate richness measurement $N_{0.5}$.

\subsection{The Hill \& Lilly Statistic $N_{0.5}$}      \label{n05}

Hill \& Lilly (1991) defined
the quantity $N_{0.5}$ to be the number of galaxies above background located 
within 0.5~Mpc radius of the quasar and with magnitudes between $m_{BCG}$ and 
$m_{BCG}$+3, where $m_{BCG}$ is the brightest cluster galaxy magnitude.  
This is similar, but not identical, to the $N_{0.5}$ used by Bahcall (1981).
To measure $N_{0.5}$ in our data, we need to know $m_{BCG}(z)$.  Also, for
valid comparison with results at low \z, we must account for redshift-dependent
changes in the bright end of the galaxy luminosity function (between $m_{BCG}$
and $m_{BCG}$+3).

\subsubsection{What is $m_{BCG}(z)$?}      \label{BCG}

Hill \& Lilly (1991) were studying the environments of radio galaxies at 
\z$\sim$0.5, and so originally defined $m_{BCG}(z)$ to be the magnitude of the 
radio galaxy in each field.  We choose to define $m_{BCG}(z)$ as the magnitude
given by the \Kz\ relation for high-\z\ powerful radio galaxies (HzPRGs),
supplemented by the \Kz\ relation of known BCGs at \z$<$1 (see \S\ref{K-z}).  
Assuming for the moment that
HzPRGs are BCGs (see below for a discussion of this point), this has the
advantage of having ``built-in'' evolutionary, cosmological, and $k$-
corrections to the BCG magnitudes.  In addition, the HzPRG population at \z=1--2
does not show much more scatter in the \Kz\ relation than reasonably expected
from studies of dozens of BCGs in clusters at \z$\leq$0.05 (\cite{eal93};
\cite{tp89}; \cite{pl95}).

One concern with estimating the BCG \Kz\ relation from that of HzPRGs is
contamination of the $K$-band light either through scattered or direct AGN
light or through AGN-induced star formation or line emission.  
Roche, Eales \& Rawling (1997) examined 10 \z=1--1.4 6C radio galaxies (RGs),
and found they have significantly smaller
$K'$-band half-light radii than 3C RGs at similar $z$.  Thus the brighter
$K$ magnitudes of the 3C galaxies are due, at least in part, to their larger
sizes rather than an increased contribution by AGN-related flux.
Best, Longair \& R\"ottgering (1997; 1998) present several other arguments in 
favor of 3C RGs being BCGs.  They claim a $\lesssim$15\% AGN-related 
contribution to the $K$-band light, and radial intensity profiles well matched 
by an $r^{1/4}$ law, both consistent with starlight from old populations
dominating at $K$.  In addition, the combined 3C RG $K$-band profile shows
evidence of excess emission at $r$$>$35~kpc, which they interpret as cD-type 
halos.  A similar excess of $K$-band emission at large radii in 3C RGs 
relative to a sample of MG radio galaxies was found by McLeod (1994), who
however suggested it may be due to very nearby (i.e. interacting) galaxies.
There is also some direct evidence that at least 
some 3C RGs at \z$>$1 lie in clusters, as expected for BCGs.
Dickinson (1997) has spectroscopically confirmed a cluster around 3C~324 at
z=1.206 from which X-ray emission has also been detected (\cite{sd95}),
and Deltorn \etal\ (1997) have spectroscopically confirmed a cluster around
3CR~184 at z=0.996.
Extended cluster-scale X-ray emission has also been detected around several 3C
RGs by Crawford \& Fabian (1996a; 1996b).

A second issue in determining $N_{0.5}$ is whether the galaxy luminosity
function (LF) down to $m_{BCG}$+3 has the same redshift evolution as $m_{BCG}$
(BCGs are about 2\fm5 brighter than $L^*$.  If not, then our values
of $N_{0.5}$ may be biased with respect to low-redshift measurements.
Best, Longair \& R\"ottgering (1998, their Figure 1) show that at \z=1--2 the
\Kz\ relation for 3C RGs runs slightly brighter than the relation for a
nonevolving ellliptical normalized to 3C RGs at \z$<$0.6.
From nearly complete redshift surveys, Cowie (1996, his Figure 2) shows that at
all \z$<$2 the upper envelope of the field galaxy population's \K\ magnitude
distribution tracks the same nonevolving ellliptical \Kz\ relation (what Cowie
plots is the \Kz\ relation of a nonevolving Sb galaxy with $M_K$=--25.8, but
the two relations are very similar at \z=1--2).
In addition, Cowie \etal\ (1996) estimate the $K$-band LF and find, within the
uncertainties, an invariant $M_K^*$ and $\alpha$ to \z=1 and consistency with
an invariant $M_K^*$ at \z=1--1.6.
These results show there must be some evolution in the BCG and field galaxy
populations since neither are ever quite as bright at \z=1--2 as the \Kz\
relation for passively evolving single-burst galaxies with $z_f$=5.
However, since both populations agree reasonably well with a single
\Kz\ relation (that of a nonevolving ellliptical), differential evolution
between them is likely to be small at \z=1--2, and so our $N_{0.5}$ values
should not be strongly biased compared to ones made at low redshift.

\subsubsection{The \Kz\ Relation for Powerful Radio Galaxies at z=1--2} \label{K-z}

To use the \Kz\ relation for HzPRGs to define $m_{BCG}$ in our quasar fields,
we must fit a function to that relation.
We use the data on HzPRGs presented in Figure 3 of Eales \etal\ (1993), which
have been corrected to 34.5~kpc apertures assuming \h=0.5, the data of
Arag\'on-Salamanca \etal\ (1993), who give $K_{CIT}$ magnitudes measured within
a 50~kpc diameter aperture for 19 BCGs in optically selected clusters at \z$<$1,
and data for the two spectroscopically-confirmed \z$>$1 BCGs (\cite{dic95};
\cite{sta97}).  Fitting the combined data, we find
\begin{equation}		 \label{eq_bcg}
K = (17.12\pm0.06) + (5.10\pm0.11) log (z)
\end{equation}	
with standard deviation 0\fm37.  
This relation is consistent with Figure 3 of Eales \etal.
We adopt Equation \ref{eq_bcg} as describing the magnitudes of BCGs to \z=2.

\subsection{Calculating the Hill \& Lilly Statistic $N_{0.5}$}   \label{calcn05}

We assume \ho=75, \qo=0.1, and $\Lambda$=0 to calculate $\theta_{0.5}$, the
angular distance corresponding to 0.5\hn1~Mpc at the quasar \z, for each quasar.
In each field we extract all galaxies inside and outside of $\theta_{0.5}$
between $K_{BCG}$ (as given by Eq.~\ref{eq_bcg}) and $K_{BCG}$+3.
However, only six fields have 5$\sigma$ limits that reach $K_{BCG}$+3.
(Our \z$>$1.4 data reaches an average of $K_{BCG}$+2\fm42, while our
\z$<$1.4 data reaches an average of $K_{BCG}$+2\fm05.)
For each other field, we scale the \no5\ measurement down to $K_{5\sigma}$
to the expected value down to $K_{BCG}$+3 using measurements to both limits 
in the six deepest fields, suitably averaged.  This eliminates the bias toward
lower \no5\ observed in the raw measurements to $K_{5\sigma}$ 
in these fields.
The uncertainties on \no5\ are dominated by the uncertainties in the background
subtraction.  There is no way to reduce the uncertainty on the number of 
galaxies within 0.5~Mpc between $K_{BCG}$ and $K_{BCG}$+3, but the uncertainty
on the background subtraction can be reduced by observing larger areas.

The results (for our liberal magnitude scale) are shown in 
Figure~\ref{fig_n05adj}.  The plotted error bars do not include the 
uncertainties in the corrections for fields which do not reach $K_{BCG}$+3.
The average \no5\ value lies above zero since we observe an excess near the
quasars.
However, the uncertainties are large:  the average \no5\ for all 31 good fields
is 11.8$\pm$12.0, for all 20 \z$>$1.4 fields is 11.4$\pm$12.8, and for all 11
\z$<$1.4 fields is 12.5$\pm$11.0.  
Using the relations between \no5, \bgq, and Abell richnesses given in 
Hill \& Lilly (1991), these \no5\ values correspond to richnesses of Abell
class 0$\pm$1, where by $-$1 we mean the richness of the general field.
In other words, the 1$\sigma$ upper limit on the average {\em near-field}
richnesses of our quasars is Abell richness class 1, not including the
uncertainty on the correction to $K_{BCG}$+3 where needed.
By {\em near-field} we mean the galaxy overdensity in the central 0.5\hn1~Mpc
radius region compared to the region beyond 0.5\hn1~Mpc in our own data.
This same Abell richness class 0$\pm$1 result was found for 17 \z=1--1.4
5C and 6C radio galaxies by Roche, Eales \& Hippelein (1997).
In Figure~\ref{fig_n05adjhist} we plot the histogram of near-field \no5\ values
for all 31 fields.  This plot can be directly compared with Figure~9 of Hill \&
Lilly (1991).  Like FR~II radio galaxies at z$\sim$0.5 (their Figure~9a) and
RLQs at \z$\sim$0.5 (their Figure~11), our quasars at \z=1--2 show values
extending from $\sim$0 to $\sim$40, with an average greater than zero
(typical Poisson uncertainties on our points are $\pm$2 bins).

We now consider the {\em far-field} richness, which we define as the galaxy
overdensity in the central 0.5\hn1~Mpc radius region compared to the literature
expectation.  We calculate these values in the same manner as the near-field
richnesses, using the
average published literature N(m) and RMS in the appropriate magnitude range.
The uncertainties are difficult to reduce other than possibly through a more
accurate determination of the random-field RMS and through observing all fields
to $K_{BCG}$+3.  The results (for our liberal magnitude scale) are shown in 
Figure~\ref{fig_n05vslit}.  These plotted error bars $do$ include the 
uncertainties in the corrections for fields which do not reach $K_{BCG}$+3.
The average far-field \no5\ is larger than the average near-field \n05,
as expected.  The uncertainties are still large:  the average far-field
\no5\ for all 31 good fields is 25.2$\pm$19.1, for all 20 \z$>$1.4 fields
is 28.8$\pm$21.4, and for all 11 \z$<$1.4 fields is 16.4$\pm$12.5.
This roughly corresponds to Abell richness 1.5$\pm$1.5 for the \z$>$1.4 
subsample, and Abell richness 0.5$\pm$1.0 for the \z$<$1.4 subsample.
Thus while the average richness levels are higher when compared to the 
literature expectation, the uncertainties are also larger due to the 
large field-to-field RMS seen in the literature.

The average ``near-field'' \no5\ value for our two control fields is
5.6$\pm$8.9, calculated assuming \z=1.201 and 1.315 for them so that their
5$\sigma$ \K\ limits are equal to $K_{BCG}$+3.  Their average ``far-field''
\no5\ value is $-$13.3$\pm$10.3 with the same assumptions.  This
negative value is expected since our control fields' galaxy counts are below 
the published literature average (Figure 9 of Paper 1).  These control field 
results show no evidence for any systematic bias in our \n05\ measurements.

\subsection{Summary and Discussion:  \n05}      \label{n05_summary}

Under the assumption that the excess galaxies are all at the quasar redshifts,
and that a straightforward comparison of \no5\ values can be done between
these quasars' galaxy excesses at \z=1--2 and clusters at \z$\sim$0, 
the values of \no5\ we calculate indicate the {\em near-field} excess
within 0.5\hn1~Mpc ($\sim$65$''$) around the quasars corresponds to Abell
richness class $\sim$0$\pm$1, where by --1 we denote the richness of the field.
This near-field excess is calculated with respect to our own data at
$>$0.5\hn1~Mpc.  We also measure the excess within 0.5\hn1~Mpc 
compared to the expected literature counts; this {\em far-field}
excess corresponds to Abell richness class $\sim$1.5$\pm$1.5.
The excess across our entire RLQ fields (i.e. to beyond 0.5\hn1~Mpc)
compared to the expected literature counts is presumably part of the same
overdensity as the far-field excess.  At some $\theta$$\gtrsim$100$''$ we expect
that this overdensity will disappear and our counts will match the literature.

Could the excess galaxies be associated with intervening \mg2\ absorption-line
systems?  Such systems would have to be at \z$\gtrsim$0.9 to explain the colors
of the excess galaxies.  From Paper 1, intervening absorber information is
available for only 4 objects at \z$<$1.4, but for 16 objects at \z$>$1.4.
There are 8 \z$>$1.4 RLQs with known intervening absorption
(14 \z$>$0.9 \mg2\ absorbers with $\overline{z}$=1.33$\pm$0.25)
and 8 \z$>$1.4 RLQs without it.
These two subsamples have $<$1$\sigma$ differences in their near- or far-field
\n05\ values despite this $\sim$3.5$\sigma$ difference in the number of
intervening absorbers along the line of sight.
Although the statistics are admittedly small, there is no evidence that
intervening \mg2\ systems contribute significantly to the galaxy excess.
This is consistent with some published evidence that \mg2\ and \civ\ absorbers
may tend to avoid clusters (\cite{be92}; \cite{mor93}; \cite{ell94}).
On the other hand, there is also evidence for excesses of galaxies at the
redshifts of strong \mg2\ and/or damped Ly$\alpha$ absorbers from narrow-band
imaging emission-line searches (\cite{bun95}; \cite{fwd97}; \cite{bec98};
\cite{man98}; \cite{tmm98}).

One simple interpretation of this result is that RLQs at \z=1--2 are located 
on a large scale ($\gtrsim$0.75\hn1~Mpc) within clusters and/or large scale 
galaxy structures of Abell richness 0 or greater (1$\sigma$ lower limit).
On a smaller scale ($\lesssim$0.5\hn1~Mpc) within those structures, the RLQs
can be located in environments ranging from the field to clusters
up to Abell richness 1 (1$\sigma$ upper limit).

However, as mentioned in \S\ref{richness}, the evolution of individual galaxies
and galaxy clusters may complicate such comparisons.  The numerical simulations
of Steinmetz (1997) show that present day $L^*$ galaxies typically have several
progenitors at \z$\sim$3 spread over a few hundred kpc, and Sawicki \& Yee
(1998) find that the inferred total stellar masses of Lyman-break-selected
\z$\gtrsim$2 galaxies in the Hubble Deep Field are not large enough for
them to be the direct progenitors of $L^*$ galaxies.  If we assume that a 
present-day $L^*$ galaxy has on average two progenitors at \z$\sim$1--2, then
our richness measurements will be biased high by a factor of two on average.  
On the other hand, early-type galaxies in clusters or other dense
structures may be fully formed at higher redshifts than field galaxies,
while clusters are probably not fully formed at high \z\ and so the galaxy
excess in our fields may need to be counted within a larger radius for
direct comparison with low redshift cluster richness measurements.
This illustrates the need for detailed comparison of high-redshift
observational data with numerical simulations in order to relate
such objects to their better understood low-redshift counterparts.
However, the above considerations will not change our basic finding that
\z=1--2 RLQs are often embedded in large-scale galaxy structures and
occasionally also in smaller structures 
the size of present-day groups or clusters.
(See \S\ref{conclude2} for a discussion of these structure's richnesses
in a broader context.)

As large as the uncertainties on \n05\ are, the near-field \n05\ values would
have had even larger uncertainties if we had not designed our data reduction
procedure to make use of the edges of the fields which have less than the full
exposure time.  More accurate determination of RLQ richnesses at \z=1--2 will
require data down to $m_{BCG}$+3 over wider fields ($>$4$'$x4$'$).
Such imaging might best be done in $J$ or $H$ if their
field-to-field galaxy RMS is smaller than in $K$.

\section{The $K$-band Luminosity Function of Candidate \z$>$1 Galaxies} \label{klf}

We have shown that an excess of red galaxies exists in our fields,
that the color distribution of the excess at \th$<$40$''$ from the
quasars is indistinguishable from that at \th$>$40$''$,
and that the color and magnitude distributions of the excess galaxies are
broadly consistent with them being galaxies at the quasar redshifts.  
Our data are too noisy for us to fully constrain the luminosity function (LF)
of the excess population under this assumption, but we can put some constraints
on $M_K^*$.  
First, some terminology:
$K_{BCG}$ is the apparent $K$ magnitude for brightest cluster galaxies (BCGs),
$M_K^{BCG}$ is the BCG absolute $K$ magnitude,
$M_K^*$ is the absolute $K$ magnitude of the knee in the luminosity function,
and $K^*$ is the apparent $K$ magnitude of the knee.

The usual method used to determine the LF is to assume a
cosmology and a prescription for $k$-corrections, use them to
determine absolute magnitudes for the excess galaxies, and fit a Schechter
function to the resulting absolute magnitude distribution.
However, since our fields reach only $\sim$2\fm5 below $K_{BCG}$ on average,
we are not sensitive to (and cannot constrain) the faint-end LF slope $\alpha$.
Similarly, since we do not know the volume over which the excess
galaxies are distributed, we cannot constrain the LF normalization $\phi^*$.
 
We assume our usual cosmology of \ho=75~\kmsm, \qo=0.1, and $\Lambda$=0.
For our no-evolution $k$-correction we adopt the average of the $K$-band
$k$-corrections for the 15~Gyr old model E and Sa galaxies of Poggianti (1997).
At the relevant rest wavelengths ($\lambda$$>$6600~\AA), these $k$-corrections
change very little over redshifts \z=1--2 and ages 11--15~Gyr.
We count all galaxies across each field brighter than the 5$\sigma$ $K$ limit
($K_{5\sigma}$), correct for incompleteness, subtract the expected literature
counts, calculate absolute magnitudes assuming the excess galaxies are all at
the quasar redshifts, sum up the individual fields' excesses, and normalize by
the area in each absolute magnitude bin.  We ignore the faintest bin
since it contains data from only one field.

The results for \z$<$1.4 and \z$>$1.4 on our liberal magnitude scale are the
solid lines in Figure~\ref{fig_klfmidzm_k} and Figure~\ref{fig_klfhizm_k}, 
respectively, with dashed lines the $\pm$1$\sigma$ Poisson uncertainty ranges.
In our cosmology our assumed $K_{BCG}$ (Eq.~\ref{eq_bcg}) corresponds
to $M_K^{BCG}$=--26.56$\pm$0.36 (including intrinsic dispersion of 0\fm3)
at \z=1.4--2, and --26.28$\pm$0.31 at \z=1--1.4.  Thus the $<$2$\sigma$ excess 
at $M_K$$<$--27, which consists of only a few observed galaxies,
is almost certainly not at the quasar redshifts.

We fit the observed excess $N$($M_K$) with a Schechter function, defined as
\begin{equation} \label{sch1}
\phi(M)=0.92 \phi^* exp (-0.92(M-M^*)(\alpha +1) - exp[-0.92(M-M^*)])
\end{equation}
(\cite{sch76}).  
Most previous $K$-band LF determinations find $\alpha$=--1.0 within their
uncertainties, and Eq.~\ref{sch1} reduces to
\begin{equation} \label{sch2}
\phi(M_K)=0.92 \phi^* exp (-exp[-0.92(K-K^*)])
\end{equation}
To estimate $K^*$, we simply assume various values of $K^*$, generate
$\phi$($M_K$) for each, and shift this curve vertically by the weighted
difference between it and all bins fainter than $M_K$=--26.5 since $\phi^*$
is unconstrained.  Objects brighter than $M_K$=--26.5 are very unlikely to be
at the quasar redshifts.  Bins fainter than $M_K$=--23.5 in the \z=1--1.4 
subsample are excluded due to their large uncertainties.  The $\chi^2$ for this
shifted curve compared to the data is calculated and and the procedure repeated
for different $K^*$ to find the value which yields the lowest $\chi^2$.
The result is $M_K^*$=--25.25$_{-\infty}^{+1.10}$ for our \z=1--1.4 fields 
and $M_K^*$=--24.85$_{-0.50}^{+0.35}$ for our \z=1.4--2 fields.
The $\pm$1$\sigma$ limits are the values of $K^*$$-$$K_{BCG}$ which yield
$\pm$1$\sigma$ deviant $\chi^2$ values.

\subsection{Discussion} \label{klf_discuss}

We compare our results with the LF of Gardner \etal\ (1997):
$M_K^*$=--23.90$\pm$0.10 at $\overline{z}$=0.14.  
Gardner \etal\ measure magnitudes in 10$''$ diameter apertures
(\cite{gar96}), or 22~kpc diameter at $\overline{z}$=0.14 for our cosmology.
This is equivalent to a 3$''$ diameter aperture at \z=1--2, smaller than the
typical total magnitude aperture for our faint galaxies, but comparable to the
typical isophotal magnitude aperture.  The average offset between these two
magnitudes is $\sim$0\fm3$\pm$0\fm2 (see Figure 4 of Paper 1).  Since our
magnitudes are probably systematically brighter than those of Gardner \etal\ by
roughly this amount, we include this offset and uncertainty in our luminosity
evolution estimates.  
Our results are consistent with mild luminosity evolution, or none:
--1.05$_{-\infty}^{+1.12}$ magnitudes in 11 RLQ fields with $\overline{z}$=1.13
and --0.65$_{-0.55}^{+0.41}$ in 20 RLQ fields with $\overline{z}$=1.67.
We stress that these estimates are made assuming that all excess galaxies in
these fields are at the quasar redshifts and with a very simple fit to an
arbitrarily normalized Schechter function with fixed faint-end slope.
Nonetheless, a somewhat more empirical estimate of luminosity evolution from
$\phi$($K$$-$$K_{BCG}$) agrees well with our estimate from $\phi$($M_K$).
In addition, this is roughly consistent
with passive evolution of stellar populations formed at high \z:
Poggianti (1997) give passive evolution of --1.05$\pm$0.10 and --1.33$\pm$0.12
magnitudes to the same redshifts for an equal mixture of Sa and E
galaxies 
in an \ho=50, \qo=0.225 universe.

If the excess galaxies were consistently located foreground to the quasars, our
estimates of their absolute magnitudes would be biased systematically bright.
To erase the evolution seen in $M_K^*$ the excess galaxies would need to have
$\overline{z}$=0.7 in the $\overline{z}$=1.13 RLQ sample, and 
$\overline{z}$=1.2$\pm$0.1 in the $\overline{z}$=1.67 RLQ sample.
This is more or less independent of the assumed cosmology.
We can rule out $\overline{z}$$\sim$0.7 since galaxies at that redshift would
not show such a strong tail to very red colors in \rK.
We cannot rule out $\overline{z}$$\sim$1.2 since galaxies can be as red as
\rK$\sim$6 at that redshift.  
Thus our detection of luminosity evolution in $M_K^*$ from $\phi$($M_K$) is
quite dependent on our assumption that the excess galaxies are all at the
quasar redshifts.  
However, if the excess galaxies are at \z$\gtrsim$0.9 as indicated by their
\rks\ color distribution, the most likely redshift for them to be at is the
quasar redshift, since there is no evidence that intervening \mg2\ systems
contribute significantly to the galaxy excess (\S\ref{n05_summary})
and the only other reason for an excess galaxy population in many fields
would be weak lensing amplification of these \z=1--2 RLQs, but large scale
structure at \z$\gtrsim$0.9 would not efficiently cause such an effect.

To summarize, in our adopted cosmology observed $K$-band BCG magnitudes match
no-evolution predictions to within 0\fm3.  
Assuming the excess galaxies are at the quasar redshifts, we estimate
--1.05$_{-\infty}^{+1.12}$ magnitudes of luminosity evolution to
$\overline{z}$=1.13 and --0.65$_{-0.55}^{+0.41}$ to $\overline{z}$=1.67.
If our assumed cosmology is in error, the amount of passive evolution we
infer will also be wrong; a \qo=0.5 cosmology would reduce our estimated
luminosity evolution by $\sim$0\fm3.

\subsection{Comparison with Previous Work} \label{klf_previous}

Our result is broadly consistent with previous work.  Studies of the $K$-band
LF through various means exist at low redshift (\cite{mse93}; \cite{gla95};
\cite{gar97}), in a \z=0.3 cluster (\cite{bea95}), to \z$\sim$1 (\cite{as93};
\cite{sdp94}; \cite{cea96}), and at \z$>$1 (\cite{as94}; see also \cite{aeo96}).
We adopted the Gardner \etal\ $M_K^*$ since it is the most accurate
value to date and lies between those of the other two low-\z\ studies.

From a $K$$<$18 field galaxy redshift survey, Elston (1994) suggested that 
$\sim$1$^m$ luminosity evolution to \z$\sim$1 in bright early-type galaxies is
required, in agreement with the results of the Lebofsky \& Eisenhardt (1986)
survey of a heterogenous sample of ellipticals and radio galaxies to \z=1.
However, Cowie \etal\ (1996) find from a $K$$<$20 field galaxy redshift survey
that within their $\pm$0\fm5 uncertainties, $M_K^*$ is invariant to \z=1 and is
consistent with being invariant at \z=1--1.6, where they observe a deficit of
bright galaxies ($M_K$$<$--25.1) despite $>$95\% complete spectroscopy,
which argues against a brighter $M_K^*$ at \z$>$1.
From a \mg2\ absorption line selected galaxy sample, Steidel, Dickinson \& 
Persson (1994) also find no evolution in the $K$-band LF to \z=1.

The studies of Barger \etal\ (1995) and Arag\'on-Salamanca \etal\ (1993) found
no evidence for evolution in $M_K^*$ (within $\pm$0\fm3--0\fm4 uncertainties)
in clusters since \z=0.3 and from \z=0.37 to \z=0.9, respectively.  
(The latter study did find evidence for color evolution of the reddest cluster
galaxies consistent with passive evolution, however.)
Also, Arag\'on-Salamanca \etal\ (1994) found that the LF of Mobasher, Sharples
\& Ellis (1993) was consistent (within uncertainties of $\pm$0\fm45)
with that of a sample of candidate \z=1.3--2 multiple \civ\ absorbing galaxies.

However, there is evidence for surface brightness evolution in both cluster and
field early-type galaxies from the work of Schade and collaborators 
(\cite{sch95}; Schade \etal\ 1996ab; \cite{slb97}; see \cite{sch96} 
for a review).  The amount of evolution at a given galaxy size is
$\Delta M_B$=--\z\ in the rest-frame $B$-band to \z=1.2.  The simplest
interpretation of the surface brightness evolution is luminosity evolution
at a level consistent with passive evolution from 
formation at high \z.  There is no evidence for differential evolution between
field and cluster early-type galaxies to \z$\sim$0.6, although the
dense cluster cores are not well represented in the cluster sample.
The results on cluster ellipticals are supported to \z$\sim$0.4
by other spectroscopic (\cite{vf96}; \cite{bzb96}) and imaging work 
(\cite{bsl96}; \cite{pdd96}; \cite{dic97}).

Thus there is no clear consensus on the magnitude of luminosity evolution in
galaxies to \z$\sim$1
(cf. Zepf 1997; Hudson \etal\ 1997; Brinchmann \etal\ 1998; Lilly \etal\ 1997).
Such evolution should be stronger in the blue, which may help reconcile the 
1\fm0$\pm$0.20 rest-frame $M_B$ evolution of Schade \etal\ with
the $<$0\fm5 limit on $M_K$ evolution of Cowie \etal.
Given the uncertainties, our results are not in conflict with previous work,
but the excess population in our RLQ fields does suggest a
brighter $M_K^*$ at \z$>$1 than at \z=0, in contrast to the trends seen 
by Cowie \etal\ (1996) and Arag\'on-Salamanca \etal\ (1994), but
in agreement with the work of Schade and collaborators.

\section{Color Pictures}	\label{colorpics}

Although color pictures themselves cannot really be used for quantitative
analysis, they are extremely useful for visualization, particularly for
comparing the colors of galaxies in different fields.
The IRAF {\sc color.rgbsun} task was used to create Sun color rasterfiles
for each field, which were converted to color PostScript using {\sc xv}.
The red, green, and blue images were the \ks, summed $r$+\ks (or $J$ when
available), and $r$ images.  The images were not convolved to the same PSF,
which leads to occasional spurious effects.  The pictures have maximum 
intensities set to a common surface brightness in mag/pixel$^2$, so that a
given color and brightness represents the same color and magnitude in every
field, except for edge effects which arise because the normalized, constant-rms
images were used to create the pictures.  

Two notable fields are shown here.  Even these high-quality plots do not
really do justice to the color images, which can be better viewed on-line at
{\it http://www.astro.utoronto.ca/$\sim$hall/thesis.html}.  In both pictures
North is at top, East at left, and the bright, blue quasar at center.
Figure \ref{fig_rjk0835} shows the field of Q~0835+580 (\z=1.534).
There is a clump of very red (\rks$\gtrsim$5) galaxies around the quasar 
(which is partially merged with a fainter star.)
Figure \ref{fig_rjk1126} shows the field of Q~1126+101 (\z=1.516).  
Of particular note are the faint orange- and red- colored galaxies to the WNW
of the quasar (bright bluish-white object at center).  These galaxies have 
similar, red \rks\ colors, but the red-colored galaxies are also quite red in
$J$$-$$K_s$, which makes them candidate background galaxies at \z$\gtrsim$2.5
(see \S\ref{sed1126}).  Also of note are the yellow-green objects $\sim$1$'$ 
ESE and $\sim$2$'$ NNE of the quasar.  These objects have red \rks\ (the ESE 
object is the reddest object in \rks\ in any field) but blue $J$$-$$K_s$, which
makes them candidate extreme late-type stars, although the NNE object may be 
extended.

\section{Spectral Energy Distributions of Candidate $z$$>$1 Galaxies} \label{seds}

In fields where we have data in filters other than just $r$ and \ks, it is
worthwhile to compare the spectral energy distributions (SEDs) of red objects 
to models of evolving stellar populations to see how strongly the galaxy
redshifts and/or stellar populations can be constrained.  Our approach is a
qualitative one since we have at most six points with which to constrain the
SEDs and since many parameters of old stellar populations are uncertain at
the $\sim$30\% level using current models (\cite{cwb96}; \cite{spi97}).

\subsection{The Candidate Group or Cluster Around Q~0835+580 (3C~205)} \label{cl0835}

There is a fairly distinct clump of objects within $\sim$20\arcsec
of Q~0835+580.  The 19 catalogued objects to our 5$\sigma$ $K$ limit
within 20\farcs5 (excluding one star at \th=20\farcs39) 
yield a surface density of $\sim$47 galaxies per arcmin$^2$.
One object 
is known to be at \z=0.236, and one 
is unresolved on the WFPC2 snapshot of this field and is likely a star.
Only one other 
shows up on the WFPC2 image, and it is definitely a galaxy.  
None of the others are likely to be stars
since WFPC2 snapshots can detect unresolved objects to very faint limits.
The compact spatial distribution of these galaxies, and the red color of many
of them, makes it very likely that they are at the redshift of the quasar 
(\z=1.534) or the intervening \mg2\ systems (\z$\sim$1.4368).

The two reddest objects at \th$<$20\farcs5 have 3\sig\ lower limits to their
\rks\ colors of 7.03 (\#339, \th=11\farcs25) and 6.79 (\#347, \th=10\farcs60;
this object is in fact an $riz$ dropout with $\gtrsim$24\fm9 in each filter,
and is detected only in $JHK_s$.)
We convert their magnitudes to $F_{\nu}$ and normalize to the same flux in \ks.
These are plotted as the filled points and error bars in 
Figure~\ref{fig_sed0835nrq2red}, for $rizJHK_s$ filters.  

To fit these SEDs,
we consider various model elliptical spectra at of the appropriate age for
the assumed \z$\sim$1.5 of the galaxies, all normalized to the data at \ks.
In our assumed cosmology the universe was 3.53~Gyr old at \z=1.534, and is
11.06~Gyr old at \z=0, slightly lower than the most recent post-Hipparcos
globular cluster age estimate of 11.5$\pm$1.3 Gyr (\cite{cha98}).  

Exponentially declining star formation histories can adequately fit ellipticals
at \z=0, but cannot fit these galaxies if they are at the quasar redshift of
\z=1.534.  This is shown by the dashed line in Figure~\ref{fig_sed0835nrq2red},
which is a 3.4~Gyr old elliptical (E) with an exponentially declining star
formation rate with $\tau$=1~Gyr from Poggianti (1997).
It fits the data at $\lambda$$>$4000$\mu$m in the rest frame, 
but is much too blue at $\lambda$$<$4000$\mu$m.  
We consider models with a shorter star formation burst for a better fit.  
Specifically, we use the Bruzual \& Charlot (1996) GISSEL96 spectral synthesis
code to produce the spectrum of a solar metallicity E viewed 3~Gyr after the
start of a 1-Gyr burst with a Scalo IMF from 0.1 to 125 $M_{\odot}$, using the
theoretical stellar spectra of Kurucz and Lejeune, Buser \& Cuisinier (1996).
However, this model (dotted line) still
produces too much flux at observed $z$ and possibly $i$ ($<$4000~\AA\ rest),
even though it fits well at $\lambda$$>$4000~\AA.  (The plotted error bars on
the fluxes do not include systematic uncertainties, and so we attach little
weight to the models' not matching the $JHK$ data simultaneously.)
As a check on the uncertainties in the stellar population modeling, we also use
the PEGASE (\cite{frv97}) code to simulate a 1-Gyr burst solar metallicity E
with Scalo IMF from 0.1 to 120 $M_{\odot}$.  Our (qualitative) results are
unchanged if we use the PEGASE model elliptical, since for this spectral range
and age it is essentially indistinguishable from the GISSEL model.
Different stellar population models can often produce very different results
(\S 7), so it is reassuring that our basic conclusions remain unchanged
whichever of these two models we use.

To produce a stronger 4000~\AA\ break to match the observations, an older
or a more metal-rich population is needed.  Dust extinction will not produce
the desired effect because the allowed reddening between 0.4--0.9$\mu$m rest
frame is smaller than the reddening needed at $<$0.4$\mu$m to match the model
with the data.  Figure~\ref{fig_sed0835nrq2red_dust} shows that a 3~Gyr
old 1-Gyr burst GISSEL model E reddened by \ebv=0.25 ($\tau_V$$\sim$0.7) 
predicts only $\sim$60\% of the observed flux at $J$.  The strength and
abruptness of the break between $z$ and $J$ (observed) requires a strong
4000~\AA\ break between those filters, either from age or metallicity.

A somewhat more speculative possible cause for the strong 4000~\AA\ break is a
nonstandard initial mass function (IMF) with a high low-mass cutoff 
$m_L$$\sim$1--5~$M_{\odot}$ (\cite{br90}; \cite{cha93}).  
Depending on $m_L$, 4000~\AA\ break amplitudes $\sim$20--40\% larger than those
of normal ellipticals can occur over a period of $\lesssim$0.5~Gyr for bursts
containing 25--100\% of the underlying galaxy mass.  
There is some evidence for high-$m_L$ IMFs in some local starbursts
(\cite{eng97}; \cite{cha93} and references therein).
Determining whether a strong 4000~\AA\ break is caused by a nonstandard IMF
rather than high metallicity would require high S/N spectroscopy.

A 6~Gyr old 1-Gyr burst GISSEL model E (solid line in
Figure~\ref{fig_sed0835nrq2red}) fits the data reasonably well at all observed
wavelengths, but this is uncomfortably old for \z=1.534.  
A $\sim$5~Gyr old model with burst duration 0.1~Gyr fits similarly well,
but 5~Gyr is still quite old for \z=1.534.
Worthey (1994) showed that two populations will appear almost identical
if $\Delta T$/$\Delta Z$=3/2, where $\Delta T$ and $\Delta Z$ are the
age and metallicity changes, in percent.  In other words, the +100\%
change in the age of the universe at \z=1.534 in our cosmology which we
require to match the observed SEDs of these two objects could instead be
produced by approximately a +67\% (+0.21 dex) change in the metallicity.
To test this, we use GISSEL96 to 
produce spectra for ellipticals formed in 1-Gyr bursts with metallicities 
$Z$=$Z_{\odot}$ ([Fe/H]=+0.0932) and $Z$=2.5$Z_{\odot}$ ([Fe/H]=+0.5595).
The results are shown in Figure~\ref{fig_sed0835nrq2red_metal}.
The high-metallicity E (solid line) is as good a fit to the data as the 6~Gyr E
model in Figure~\ref{fig_sed0835nrq2red}.  
Thus the SEDs of these objects are consistent with $\sim$3~Gyr old metal-rich 
ellipticals, even though the metallicities of \z=0 ellipticals averaged over
the entire galaxy like our SEDs are likely to be roughly solar (\cite{spi97}
and references therein).

However, these two objects are $redder$ than LBDS 53W091 (\z=1.552), a mJy
radio source with \rk$\sim$6.15 (\cite{dea96}) whose ultraviolet spectrum is
remarkably similar to M32.  M32 is believed to contain an intermediate-age
stellar population $\sim$4--5~Gyr old in addition to the old stellar population
usually present in ellipticals (\cite{spi97}, and references therein).  Spinrad
\etal\ (1997) claim an age of $\gtrsim$3.5~Gyr for LBDS 53W091, and Dunlop
(1996) quote $\gtrsim$4.5~Gyr for a similar galaxy (LBDS 53W069 at \z=1.432).  
On the other hand, Bruzual \& Magris (1997) claim that 
LBDS 53W091 can be modelled as an elliptical only 1--2 Gyr old
if the full rest-frame 0.2--1$\mu$m SED is taken into account,
The discrepancy is partially due to the evolutionary spectral synthesis models
they use (\cite{bc93}), which produce red rest-frame $R$$-$$K$ colors as fast
or faster than any other such models (\cite{spi97}, Figure~16).
However, it is also partially due to 
the rest-frame optical-IR colors of LBDS 53W091 being 
bluer than those of M32 despite the similarities in their rest-frame UV spectra.
(Heap \etal\ (1998) also claim an age of $\sim$2~Gyr for LBDS 53W091, on the
independent basis of comparison with a Space Telescope Imaging Spectrograph
ultraviolet echelle spectrogram of an F8V star.)

In Figure~\ref{fig_sed0835nrq2redvsm32} we compare the SEDs of our two 
\rks$\sim$7 objects to the spectrum of M32 (\cite{bic96}), normalized at \ks.  
Within the systematic uncertainties, the M32 spectrum is a good fit.  
This is consistent with the good fit of the 6~Gyr model in 
Figure~\ref{fig_sed0835nrq2red}.  
The Bica \etal\ M32 agrees well with the 0.4--0.65$\mu$m spectrum of Hardy \&
Delisle (1996), but is bluer than the ``M32(b)" SED of Magris \& Bruzual (1993).
Even adopting the latter spectrum, however, the main objection of Bruzual \&
Magris (1997) to the old age claimed for LBDS 53W091 at \z$\sim$1.5 is
substantially weaker for these two galaxies at similar \z, and possibly
for a third object with \rks$\gtrsim$6.5 in the field (\#618, \th=99$''$).
Deep optical spectra of these objects have the potential of placing constraints
on the cosmological model as strong as or stronger than those claimed for LBDS
53W091 by Spinrad \etal\ (1997).

The seven objects next reddest in \rks\ at \th$<$20\farcs5 have \rks=4.95--5.80.
The remaining nine at \th$<$20\farcs5 --- excluding the galaxy of known
\z=0.236 --- have \rks=2.68--4.49, consistent with any \z$\gtrsim$0.5,
and we do not consider them further.  We normalize these objects to the
same flux in \ks and compute the weighted average $F_{\nu}$ in each filter.
We plot the average and RMS scatter in Figure~\ref{fig_sed0835nrq7red} and
compare to model spectra.
The 3.4 Gyr old Poggianti E model produces too much flux in observed $riz$.
A 2 or 3~Gyr old GISSEL model is a decent fit:  the 3~Gyr old model (dotted
line) fits best in $z$, and the 2~Gyr old model (solid line) fits best in $i$,
but both predict too little flux in $r$.  This can be interpreted to mean
that, on average, star formation dropped off in these objects more quickly
than an exponential but less quickly than a sudden cutoff of a burst.  
The inferred ages of these very red objects within 20$''$ of 
3C~205, and others at larger distances from the quasar, do not strongly
constrain cosmological models.  However, if they truly are younger than the
extremely red objects (EROs) with \rks$\gtrsim$6.5 discussed earlier and yet
are also early-type galaxies at the same redshift, it would mean that
\z$\simeq$1.5 is near enough to the formation epoch of these galaxies
for the dispersion in the rest frame $U$$-$$V$ colors of early-type
galaxies to rise above the small values seen at \z$\lesssim$1 (\cite{sed98}).

\subsection{The Southern Radio Hotspot of 3C~205}	\label{hotspot}

Lonsdale \& Barthel (1986; 1988; 1998) have studied the radio morphology of
this quasar at VLBI resolution.  The southern hotspot is extremely compact, 
curved, and polarized, and is is located 2\farcs3 W and 8\farcs2 S of the
quasar.  One of the extremely red galaxies discussed above (\#347) is located
3\farcs80 W and 9\farcs89 S of the quasar ($\theta$=10\farcs60), consistent
with the hotspot being produced by the collison of the radio jet with part of 
this galaxy.  (There may be a small ($<$3$\arcdeg$) rotation of our image
relative to true north.)  This ``hotspot galaxy'' has \ks=19.53$\pm$0.07,
\rks$>$6.79 (3$\sigma$ lower limit), $z$$-$$J$$>$3.89 (3$\sigma$),
\jks=1.87$\pm$0.06, and \hks=0.85$\pm$0.18.
As we saw in \S\ref{cl0835}, this object's SED is red enough to match that of a
metal-rich and/or very old galaxy, but its observed $>$1$\mu$m slope does
not allow much dust extinction.  Thus the jet is probably being deflected by a
high-pressure X-ray halo around the galaxy rather than dusty gas within it.

Could the hotspot object be synchrotron emission from the hotspot itself?
The SED of such emission is expected to have $f_{\nu} \propto \nu^{-1}$
(\cite{rs97}), or \rks=3.33, $z$$-$$J$=1.49, \jks=1.61, and \hks=0.81.
The near-IR colors of the hotspot object are consistent with the expected
synchrotron emission, but such emission cannot explain the break at
$\sim$1$\mu$m, observed, which is most naturally explained as a 4000~\AA\ break
at \z$\sim$1.5, consistent with the quasar redshift.

\subsection{Candidate Associated and Background Galaxies in the Field of Q~1126+101} \label{sed1126}

A very intriguing feature of the field of Q~1126+101 is that many of the
\rks$>$4 galaxies have very red \jks\ colors as well (cf. the reddest objects in
Figure~\ref{fig_rjk1126}).  
In this section we show that the SEDs of these objects are most consistent with
either galaxies at \z$\gtrsim$2.5 ({\em background} to the quasar) whose
4000~\AA\ breaks lie between $J$ and $K_s$,
or with very dusty galaxies at the quasar redshift of \z=1.516.  
Many 
of these objects have \rks\ and \jks\ colors similar to the spectroscopically
confirmed \z=2.38 galaxies of Francis, Woodgate \& Danks (1997), and 
several are actually redder in $J$$-$$K$ than the only ERO with a spectroscopic
redshift, HR10 (\cite{hr94}), which requires dust reddening in any reasonable
cosmology to fit its SED at its known \z=1.44 (\cite{gd96}; \cite{cim97}).

We attempt only qualitative SED modelling, first trying 
to fit each object with elliptical spectra of reasonable age
for the assumed $z$ to minimize the required dust extinction.
To model the effects of dust, we used the standard Milky Way (MW), LMC, and SMC
extinction laws and the Calzetti (1997) law.
The Calzetti formula was derived for active star-formation regions and 
empirically incorporates the ``selective attenuation'' effects of dust,
including extinction, scattering, and geometrical dust distribution effects.  
In any case, the relative
extinction from $\sim$2500 to 9000~\AA\ (observed $r$ to $K$ at \z=1.5) is
similar for all four of these curves, as seen in Figure~\ref{fig_sedust1126b}.

In Figure~\ref{fig_sedust1126b} we plot the two objects with \rks$>$5 which are
reddest in \jks\ (\#424 and \#315).  A 2.2~Gyr old $\tau$=1~Gyr Poggianti E at
\z=2.5 with \ebv=0.35 ($\tau_V$$\sim$1) provides a good match to the SEDs of
these objects.  The choice of reddening law does not strongly affect the fit
except at $r$ 
and possibly at $z$.  
The break between $J$ and
$H$ is suggestive of \z$\sim$2.5 for these two objects, but the lack of detected
flux except at $r$ and \ks\ makes for a weak constraint on the redshift.  
However, a lower redshift would require a larger $\tau_V$ since longer
rest-frame wavelengths would be observed, and a younger galaxy 
or one with a more extended star-formation history would require more dust.
The required reddening can be reduced to \ebv$\sim$0.25 by fitting with a
$\sim$2-Gyr old 1-Gyr-burst and a young ($\lesssim$0.1~Gyr) population
contributing substantial light only at $<$1$\mu$m (observed).  
Similar \ebv\ values were found by Sawicki \& Yee (1998) for confirmed
\z$>$2 galaxies in the Hubble Deep Field (HDF) through detailed SED fits.

Figure~\ref{fig_sedust1126c} 
shows that the same \ebv=0.35 Poggianti E fits the $J$ and $K_s$ fluxes of the
two objects with \rks$<$5 which are reddest in \jks\ (\#18 and \#237).
However, these objects definitely require an additional young population to
match the blue $r$$-$$J$ colors.  

The lack of usefully deep $H$ data for the above
objects prevents firm redshift estimates.
Our assumed \z=2.5 is the lowest 
that places the 4000~\AA\ break redward of
$J$, but while strong breaks in the objects' spectra are consistent with the
observed SEDs, they are not necessarily required.
Figure~\ref{fig_sedust1126d} shows the 
only two objects with \jks$>$2.5 with firm detections in $H$
which might help better constrain their redshifts (\#425 and \#381).
These objects are more consistent with dusty galaxies at the quasar redshift
\z=1.516 than objects at \z$\sim$2.5 due to the strong spectral break between
$z$ and $J$ (compare to the objects in Figures~\ref{fig_sedust1126b} and
\ref{fig_sedust1126c} which cannot have such strong breaks).
Such objects are unusual; there are only two other \jks$>$2.5 galaxies which
definitely have similarly strong breaks.  
The red \jks\ and blue $r$$-$$z$ colors can be fit only with composite stellar
populations, e.g. a dusty 2~Gyr old population with \ebv=0.75 ($\tau_V$$\sim$2)
and a young (100~Myr) dust-free population with $\sim$1\% of the total mass.
The only way to avoid dust would be with metallicities $Z$$>$5$Z_{\odot}$
and/or ages $\gtrsim$5~Gyr, plus a young component providing flux in $r$.

\subsection{Candidate Dusty Associated and/or Background Galaxies in Other Fields} \label{otherzg2}

The presence of candidate dusty galaxies around RLQs at \z$\sim$1.5 is relevant
to the evolution of early-type galaxies in clusters and groups, and the surface
density of candidate \z$\gtrsim$2.5 galaxies is interesting in its own right,
so we briefly examine \jks$>$2.5 objects in the other fields 
with deep $J$ data (Q~0835+580, Q~0952+179 and Q~1258+404).  

The surface densities of objects with \jks$>$2.5 and \K$<$21 are
4.31~arcmin$^{-2}$ in the Q~1126+101 field and 1.90~arcmin$^{-2}$
in the other three fields.  
If we also require $r$$<$25.5 the surface densities are
2.80 and 1.29~arcmin$^{-2}$ respectively, to $\pm$25\% accuracy.
Whatever these galaxies are, the Q~1126+101 field is clearly unusual.

It is worthwhile to compare the surface density of these candidate
\z$\gtrsim$2.5 ``$J$-dropouts" (objects with \jk$>$2.5)
with that of ``$U$-dropouts" at \z$\sim$3.0$\pm$0.5.
From Steidel \etal\ (1998b), to $R_{AB}$=25 (or $r$=25.5) there is
$\sim$1 $U$-dropout/arcmin$^{2}$.  The four $U$-dropouts with published
$K$ data have $K$=21.3--22.1 and $R_{AB}$$-$$K$=2.7$\pm$0.5 (\cite{ste96}).
Thus, at the same $r$ magnitude the surface densities of $J$- and $U$-dropouts
are approximately equal, but our $J$-dropouts are about $\sim$2$^m$ brighter
in $K$ and thus $\sim$2$^m$ redder in \rk\ than $U$-dropouts.  This implies
that many $J$-dropouts should be found among $U$-dropouts studied in the IR.
Only four with \jk$>$2.5 have been seen among the four objects mentioned above
and the 17 $U_{300}$-dropouts in the HDF studied by Sawicki \& Yee (1998).  
These results could be reconciled if faint \z$\sim$2.5 galaxies are on average
redder than faint \z$\sim$3 galaxies, but if anything the opposite is seen by
Sawicki \& Yee (1998).
To reduce the discrepancy between the inferred surface densities of
$J$-dropouts at \z$\gtrsim$2.5 and $U$-dropouts at \z$\sim$3, 
we are led to the conclusion that a substantial fraction of our $J$-dropouts
are not at \z$\sim$2.5 but are dusty galaxies at the quasar redshifts.
Even then, the relative surface densities are such that we expect up to 50\%
of $U$-dropouts will be $J$-dropouts, larger than the 20$\pm$10\% seen to date
among the 21 $U$-dropouts with published near-IR data.

This conclusion is consistent with the surface densities of $J$-dropouts in
literature control fields with $J$ data (Appendix~\ref{litcf}).  
In 43.14 arcmin$^2$ to \K=21 the surface density of
\jks$>$2.5 galaxies is 0.46~arcmin$^{-2}$, 
about half that in our fields (excluding Q~1126+101).  This suggests that
roughly half our $J$-dropouts are in fact associated with the quasars.

The SEDs of our $J$-dropouts are also consistent with about half
being dusty galaxies at the quasar redshifts $z_q$.
Only Q~0835+580 and Q~1126+101 have $z$- and $J$-band data which is needed
to identify the 4000\AA\ break at $z_q$$\sim$1.5.
If we select \ks$<$21 objects with \jks$>$2.5 and \rks$>$5
(indicating a possible break between $r$ and \ks), 7 of 22 such objects in the
Q~1126+101 field and 7 of 10 in the Q~0835+580 field (roughly half in all)
have $z$$-$$J$ colors or lower limits consistent with dusty galaxies at $z_q$.

\subsection{Summary:  Spectral Energy Distributions} \label{seds_summary}

There is a distinct overdensity of galaxies around Q~0835+580 (3C~205).  
The extremely red spectral energy distributions (SEDs) of the two reddest
objects within $\sim$20$''$ (and a third object 99$''$ away) could be due to
metallicities $\gtrsim$+0.2~dex above solar or to stellar populations as old as
M32 (4--5~Gyr) at \z$\sim$1.5.
Dust extinction cannot explain the entire SEDs due to the strong break
between $z$ and $J$ (observed), presumed to be the 4000~\AA\ break.
The SEDs of the seven next reddest objects in \rks\ within $\sim$20$''$ of
3C~205, and of many other galaxies across that field, are consistent with
2--3~Gyr old ellipticals formed in a 1-Gyr burst.  
If these galaxies and the three even redder ones are all early-type galaxies
at the quasar redshift, then we may be seeing a dispersion in the colors,
metallicities, or ages of early-type cluster galaxies or their progenitors.

In the field of Q~1126+101, there are three dozen objects with \jks$>$2.5,
a distinct excess compared to other fields, which themselves show an excess
of such ``$J$-dropouts" compared with literature control fields.
Several lines of argument suggest about half of these galaxies are most likely
old and dusty ($\sim$2~Gyr, \ebv$\sim$0.75) galaxies at the quasar redshifts
with young ($<$0.1~Gyr) components comprising $\sim$1\% of their masses.
The remaining $J$-dropouts are consistent with being at \z$\gtrsim$2.5,
which places the 4000~\AA\ break between $J$ and \ks.

The possible presence of relatively old, dusty galaxies at the quasar redshifts
coupled with the possible dispersion in early-type galaxy ages mentioned earlier
suggests there may be considerable dispersion in the properties of cluster
ellipticals at \z$\sim$1.5, in agreement with hierarchical clustering models
(\cite{kau96}).  However, these candidate relatively old, dusty galaxies could
also be younger systems with higher \ebv.

\section{Conclusions}	\label{conclude2}

The major results of our study so far are:  

\noindent{$\bullet$~We find a significant excess of $K$$\gtrsim$19 galaxies in
the fields of 31 $z$=1--2 RLQs, on two spatial scales (\S\ref{radprof}).
One component is at $\theta$$<$40$''$ from the quasars and is significant
compared to the galaxy surface density at $\theta$$>$40$''$ in the same fields.
The other component appears roughly constant across the fields to
$\theta$$\sim$100$''$ from the quasars and is significant compared
to the galaxy surface density seen in random-field surveys in the literature.
The $\theta$$<$40$''$ component may be produced by as few as $\sim$25\% of the
\z$>$1.4 fields but the large-scale component is present in $\gtrsim$50\% of
them.}

\noindent{$\bullet$~The $r$$-$$K$ color distributions of the excess galaxy
populations are indistinguishable from each other and are significantly redder
than the color distribution of the field population, consistent with the excess
galaxies being predominantly at \z$>$1 (\S\ref{cc1d}).
However, there is a deficit of blue (\rK$\lesssim$3.5) galaxies at \k=20--21
which is difficult to understand as either real or spurious.}

\noindent{$\bullet$~The magnitudes and colors of the excess galaxies are
consistent with a population of mostly early-type galaxies at the quasar
redshifts, such as would be found in quasar host clusters or groups.  
There is no evidence that they are associated with intervening \mg2\ absorbers
(\S\ref{n05_summary}).}

\noindent{$\bullet$~The average excess within 0.5$h_{75}^{-1}$~Mpc 
($\sim$65$''$) of the quasars corresponds to Abell richness class $\sim$0$\pm$1
compared to the galaxy surface density at $>$0.5$h_{75}^{-1}$~Mpc from the
quasars, where --1 denotes the richness of the field,
and to Abell richness class $\sim$1.5$\pm$1.5 compared to the galaxy
surface density from the literature (\S\ref{richness}).
This assumes that the excess galaxies are all at the quasar redshifts.
This suggests that on a large scale ($\gtrsim$0.75\hn1~Mpc) RLQs at \z=1--2 are
located within clusters and/or large scale galaxy structures of Abell richness
$\sim$1.5.  On a smaller scale ($\lesssim$0.5\hn1~Mpc) within those structures,
RLQs can be located in unremarkable ``field'' environments or in groups or
clusters of Abell richness $\sim$0.
One uncertainty not taken into account when calculating
richness measurements is the correspondence between individual galaxies at
\z$>$1 and \z$\sim$0.  A high merger rate at \z$<$1--2 would mean that
individual galaxies at \z$\sim$0 were composed of several progenitors at
\z$>$1--2, which could bias our richness measurements high.
Spectroscopy and detailed comparison with numerical simulations are needed
to resolve this uncertainty.}

\noindent{$\bullet$~The only significant correlations observed between
environments and quasar properties is that the galaxy excess at
$\theta$$<$40$''$ seems to be stronger for the more radio-powerful and
steeper-radio-spectrum RLQs (\S\ref{radprof_correlate}).  The $\theta$$<$40$''$
excess does not seem to depend on the presence of associated absorption.
These dependences are based on only 31 RLQs (or fewer), but they at least
illustrate trends which could be verified with larger datasets.}

\noindent{$\bullet$~By assuming the excess galaxies are at the quasar redshifts
and fitting their $K$-band luminosity function, we find --0.65$_{-0.55}^{+0.41}$
magnitudes of luminosity evolution in $M_K^*$ to $\overline{z}$=1.67
(\S\ref{klf}).  This is in contrast to the trends seen at \z$>$1 
by Cowie \etal\ (1996) and Arag\'on-Salamanca \etal\ (1994), but plausibly 
in agreement with the work of Schade and collaborators (\cite{sch96}).}

\noindent{$\bullet$~For four fields with data in at least $rJK_s$, we find that
the SEDs of most of the excess galaxies are consistent with them being 2--3~Gyr
old early-type galaxies at the quasar redshifts of $z$$\sim$1.5, but that there
are galaxies whose SEDs cannot be fit by such simple models (\S\ref{seds}).  
At least three objects in these four fields have SEDs consistent with being
4--5~Gyr old at $z$$\sim$1.5, and at least a dozen others are consistent with
$\sim$2~Gyr old but dust-reddened galaxies at the quasar redshifts
(\S\ref{otherzg2}).
Taken together, these potentially different galaxy types suggest that there is
considerable dispersion in the properties of early-type cluster galaxies at
$z$$\sim$1.5, in agreement with hierarchical clustering models (\cite{kau96}).
However, these galaxies could also be younger systems with higher \ebv.
Spectroscopic followup will be needed to confirm this suggestion.
In particular, age determinations from deep spectra of the candidate 4--5~Gyr
old galaxies offer the possibility of constraining the cosmological model by
requiring a relatively old universe at large lookback times.}

\noindent{$\bullet$~There are also several dozen galaxies in the four fields
with good $J$ data (particularly in the Q~1126+101 field) whose SEDs are best
explained either as background galaxies at $z$$\gtrsim$2.5 or very dusty
galaxies at \z=$z_q$ (\S\ref{sed1126}).  Some of the $z$$\gtrsim$2.5 candidates
seem to be dusty, to have composite stellar populations,
or both, and some may be already $\gtrsim$2~Gyr old at $z$$\gtrsim$2.5,
again offering a possibile constraint on cosmological models.}

Without spectroscopic confirmation it is premature to draw firm conclusions
about the implications of the existence of the excess galaxy population in
these \z=1--2 RLQ fields.  Nonetheless, if we assume that the excess galaxies
are predominantly at the quasar redshifts, we can draw some interesting
conclusions which illustrate the value of spectroscopic followup.

The existence of galaxy excesses on two spatial scales, consistent with Abell
richness $\sim$0$\pm$1 clusters embedded in Abell richness $\sim$1.5$\pm$1.5
structures if the
excess galaxies are at the quasar redshifts, might lend support to hierarchical
clustering models independent of $\Omega$.  This is because in such models
clusters are built up by accretion of smaller clusters and groups over time.
Galaxy structures embedded in larger structures such as we observe could occur
during some phases of cluster formation in hierarchical clustering models.

This hierarchical clustering interpretation of the galaxy excesses on two
spatial scales may tie in with RLQ host cluster observations at \z$<$0.7
(see \S\ref{intro} and the introduction to Paper 1).
At such redshifts the clusters appear to be younger 
and less virialized than optically-selected clusters (\cite{egy91}) 
and are of Abell richness 0--1, with a few of richness 2.
At \z=1--2 we also find that RLQs can be found in overdensities of
Abell richness 0--1, but that these ``near-field'' overdensities
are embedded in larger structures of Abell richness $\gtrsim$1.  
It is an open question
whether the most important factor affecting the evolution of RLQs in clusters
is the absolute density of the environment, the density within 0.5~Mpc,
or even conditions in the host galaxy and any galaxies interacting with it.
However,
despite the fact that an Abell richness 0 (sub)cluster observed at \z=1.5 will
evolve into a very different structure at \z=0 than an Abell richness 0 cluster
observed at \z=0.5 will, we observe RLQs in Abell richness $\sim$0 
(within 0.5$h_{75}^{-1}$~Mpc) clusters at both \z$\sim$0.5 and \z$\sim$1.5.  
This may imply that it is the properties of a RLQ's
environs within 0.5~Mpc which determine whether the environs
are hospitable to RLQ formation and fueling; whether or not the Abell richness
$\sim$0 cluster is embedded in a larger structure or not may be immaterial.
Alternatively, if clusters embedded in different large-scale environments show
important differences in their formation histories, then the large-scale
density of the RLQ's environs would be what determines which clusters host RLQs.
If this is the case then the important result is that quasar host
clusters at \z$\sim$1.5 are richer on large scales than at \z$\sim$0.5.

If we assume instead that the interpretation of Ellingson, Green \& Yee (1991)
is correct, i.e. that galaxy interactions are largely responsible for creating
and fueling RLQs and that only dynamically young clusters are hospitable to RLQ,
then one interpretation of our \z=1--2 results would be that we are seeing
dynamically young subclusters which will later
virialize and/or merge with other galaxies or subclusters in the observed
large-scale galaxy structures to form clusters as we know them today.  
The fact that RLQ host clusters at \z=1--2
are not markedly richer on 0.5$h_{75}^{-1}$~Mpc scales
than those at \z$<$0.7 is again consistent with hierarchical clustering models.
In such models richer clusters do not undergo monolithic collapse at higher \z,
but instead form from mergers of subclusters at higher \z.

The above speculations should be treated as such.  
Spectroscopic investigation of these fields 
and detailed comparison to numerical simulations is needed before we can begin
to formulate a true model of the evolution of quasar environments to \z=2.  
Nonetheless, if and when such confirmation is obtained, we feel that the above
discussion outlines a reasonable framework for future thinking on this subject.

Among the followup research projects suggested by this work, two will be
very important in confirming (or denying) and extending our results.
First, larger areas ($>$4$'$x4$'$) around our sample need to be imaged to
confirm the large-scale ($\theta$$\sim$100$''$) galaxy excess around
\z=1--2 RLQs and to determine the true angular extent of this excess.  Second, 
multislit optical and near-IR spectroscopy 
is needed to verify the existence of overdensities at the quasar redshifts,
to determine velocity dispersions of any large-scale structures found, and
to discriminate age, metallicity, and dust effects in the brightest candidate
\z$>$1 galaxies.  Over the next decade IR spectrographs will begin to appear
on 8-m class telescopes (e.g., the MMT 6.5m, LBT, and Keck) and perhaps
the Next Generation Space Telescope.  These instruments will enable
spectroscopic identification of very red and faint galaxies and 
studies of their ages, metallicities, and dust reddenings.  Ultimately, they
may enable the fundamental plane of elliptical galaxies and its evolution to
be studied to \z=1--2 and beyond using galaxies from this and other work.

\acknowledgements

This work was part of a Ph.D. thesis at the University of Arizona.
PBH acknowledges support from an NSF Graduate Fellowship and from NASA.
We thank M. Dickinson and R. Elston and their collaborators for use of data
prior to publication, all authors who have made all or part of their $K$
band imaging datasets publicly available for ourselves and others to use,
and the referee for interesting and helpful comments.
This research has made use of observations made at the Kitt Peak National 
Observatory, National Optical Astronomy Observatories, which is operated by
the Association of Universities for Research in Astronomy (AURA), Inc., 
under contract to the National Science Foundation,
and at the Infra-Red Telescope Facility, which is operated by the University
of Hawaii under contract to the National Aeronautics and Space Administration;
the NASA/IPAC Extragalactic Database (NED), operated by the Jet Propulsion
Laboratory, California Institute of Technology, under contract to NASA;
and data from operations made with the NASA/ESA Hubble Space Telescope, 
obtained from the data archive at the Space Telescope Science Institute, 
which is operated by AURA, Inc., under NASA contract NAS 5-26555.

\appendix
\section{Literature Control-Field Data}      \label{litcf}

Since our own control fields cover far less area than our RLQ fields,
it is advantageous to have as much additional control field data as possible,
preferably with data in $r$ and \ks.  Of the datasets available in the
literature, however, few utilize those specific filters.
For purposes of comparing color-magnitude diagrams, it is acceptable to use
data in various optical filters to estimate $r$ magnitudes, since the 
uncertainties in predicting \rks\ from $I$$-$$K$ (e.g.) for faint galaxies are 
smaller than the binning size of 0\fm5 we will typically use in our comparisons.
To estimate galaxy colors in one optical filter set from another, we use the
galaxy colors for morphological types E through Im which have been computed by
Frei \& Gunn (1994; FG94) from z=0--0.6 using the SEDs of Coleman, Wu \& Weedman
(1980), and by Fukugita, Shimasaku, \& Ichikawa (1995; FSI95) from z=0--0.8
using the spectral atlas of Kennicutt (1992).
Specifically, we use Tables 2--6 of FG94 and Tables 3--9 of FSI95 where both
references give consistent first-order color-color conversion results (after 
accounting for different zeropoints in the two references, cf. Appendix A of
Paper 1) and FSI95 alone where they do not.

At \ks$<$20, for \rks\ comparisons we use primarily 
our two control fields (18.78 arcmin$^2$, $R_C$ and \ks), 
the McLeod \etal\ (1995) Her-1 field (10.56 arcmin$^2$, Tyson $R_T$ and \ks),
the HDS (\cite{cow94}) SSA13, 17, and 22 fields (5.33 arcmin$^2$, $U'BVI_C$
and $K$), and preliminary data (25 arcmin$^2$, $R_C$ and \ks)
from the SA57 field of the 100 arcmin$^2$ $BRIzJK$ KPNO 4m+IRIM survey of 
Elston, Eisenhardt, \& Stanford (1998; EES98), 
kindly provided by R. Elston.  At \ks$>$20, we use
the Djorgovski \etal\ (1995) Hercules field (0.44 arcmin$^2$, $r$ and \ks),
the Moustakas \etal\ (1997) Fields I and II (2.88 arcmin$^2$, $VIK$), and 
the preliminary KPNO 4m+IRIM catalogs of the Hubble Deep Field (HDF-IRIM; 
7.40 arcmin$^2$, $F606W,F814W,JHK$), kindly provided by M. Dickinson
(cf. \cite{dic98}).
Conservative uncertainties of $\pm$0\fm1 were assigned to magnitudes, and $\pm$0\fm15 
to colors, where none were given.  We only removed objects classified stellar
by the various authors, who used varying criteria including morphologies and
colors (cf. Saracco \etal\ 1997).  Stellar contamination is not a large effect
at faint magnitudes, but should be kept in mind nonetheless, particularly for 
the HDF-IRIM and EES98 datasets which still include stars.

We estimated the $r$ and $i$ magnitudes of the McLeod objects from their $R_T$ 
and $I_T$ magnitudes (\cite{gul95}) using relations derived from FSI95 and
our adopted zeropoints (see Appendix A of Paper 1 and Appendix~\ref{litcf}
of this paper):
\begin{eqnarray}
r=R_T+0.367-0.048\times(R_T-I_T)\\ 
i=I_T+0.710+0.045\times(R_T-I_T) 
\end{eqnarray} \label{eq_ri_from_RIT}
We estimated the $r$ magnitudes of the EES98 objects from
their $R_C$ magnitudes using a relation derived from both FG94 and FSI95:
\begin{equation}
r=R_C+0.322 
\end{equation} \label{eq_r_from_RC}
This is the same relation used for our own control fields and the Q~2230+114 
field, and the color term is negligibly small.
We estimated the $r$ magnitudes of the Moustakas and HDS objects from their 
$V$ and $I_C$ magnitudes using a relation derived from both FG94 and FSI95:
\begin{equation}
r=I_C+0.447+0.452 \times (V-I_C).  
\end{equation} \label{eq_r_from_VI}
The 1$\sigma$ scatter around this fit is the largest of all our fits, but is
still $\pm$0\fm1 or less at z$\lesssim$1.
(The published $V$ and $I$ magnitudes for Field II of Moustakas \etal\ are 
in error and should be adjusted by $-$1\fm6 and +1\fm6 respectively.)
For the Dickinson \etal\ (1998) data, we convert the HDF $F606W$ and $F814W$ AB
magnitudes to Vega-based magnitudes which we dub $V_{606}$ and $I_{814}$ using 
$V_{606}$=$F606W$$-$0.116 
and $I_{814}$=$F814W$$-$0.439 (\cite{wil96}).
We then convert to $V$ and $I_C$ using $V$=$V_{606}$+0.37($V_{606}$$-$$I_{814}$)
and $I_C$=$I_{814}$$-$0.10($V_{606}$$-$$I_{814}$), which are first-order 
approximations derived from Tables 10 and 7 of Holtzman \etal\ (1995).
Finally, we estimate $r$ from $V$ and $I_C$ as for the Moustakas data above.  

Control field $J$ band data were taken directly from the McLeod \etal\ (1995)
SA57SO field, the EES98 SA57 field, and the HDF-IRIM data (\cite{dic98}).

We use several different combinations of these datasets in our work.
We exclude the Soifer \etal\ (1994) data 
since those fields were targeted around high redshift objects and are thus not
the random fields we desire as control fields.
By ``published control fields'' we refer to all datasets mentioned above,
except the HDF-IRIM and EES98 datasets, plus the bright-end data of Glazebrook
\etal\ (1994) and the HMWS and HMWS (\cite{gar95a}; \cite{gar95b}).
By ``opt-IR control fields'' we refer to all datasets from which
we generated \rks\ colors, including our own.
This includes only parts of some datasets:  all but the HDS SSA4 field of
Cowie \etal\ (1994), just the Hercules field of Djorgovski \etal\ (1995), 
just the Her-1 field of McLeod \etal\ (1995), and just the SA57 field of EES98.

\begin{figure}
\plotone{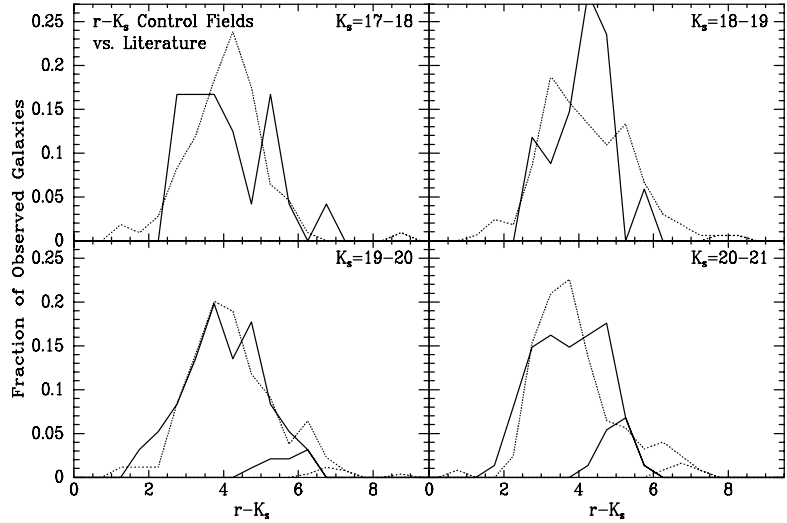}          
\caption[$r$$-$$K_s$ Galaxy Color Histograms: All Literature Data vs. $r$$-$$K_s$ Literature Data]{
\singlespace
Comparison of $K_s$-selected galaxy \rks\ colors between fields with actual
$r$ (or $R$) and $K$ data (solid lines) and literature control fields with $r$
and $K$ data synthesized from magnitudes in other filters (dotted lines).  
Smaller histograms are galaxies with lower or upper limits to their colors.  
Each histogram has been separated normalized to unity sum.
The $K_s$ magnitude range is given in each panel.
}\label{fig_bin1dnorm_cfrkvslit}
\end{figure}

To check 
that our conversions of the various literature colors to \rk\ colors are
accurate, we compare \rk\ distributions in Figure \ref{fig_bin1dnorm_cfrkvslit}.
The dotted lines are all opt-IR control fields except our own, and the solid
lines are our own control fields and those of Djorgovski \etal\ (1995), which
are the only fields with data originally taken in $r$ (or $R$) and $K_s$.
The uncertainties are large, but there are no significant differences between
the actual and converted \rk\ distributions at any \K=17--21,
i.e. between our control field data and that from the literature,
given the deeper limits of many of the opt-IR control fields.


\clearpage

\clearpage

\begin{figure}
\plotone{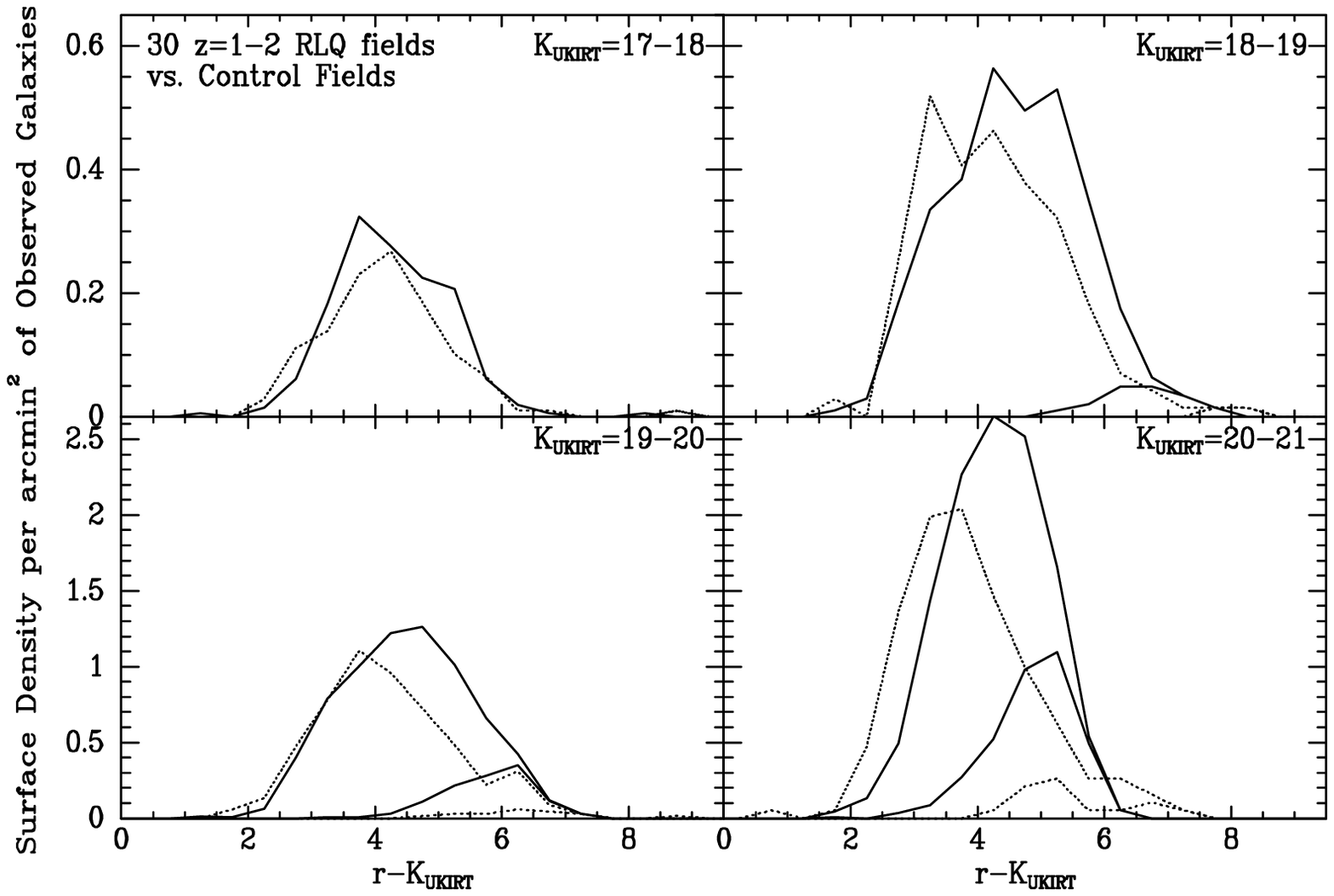}      
\caption[$r$$-$$K$ Galaxy Color Histograms in $z$=1--2 RLQ Fields: Surface Density of Observed Galaxies]{
\singlespace
Colors of $K$-selected galaxies.
Solid line is \z=1--2 quasar fields; dotted line is control fields.  Smaller
histograms represent those galaxies with lower or upper limits to their colors.
Each histogram has been normalized by the area imaged in the appropriate
magnitude range to yield the surface density of galaxies per arcmin$^2$ 
in each color bin.  The \Kuk\ magnitude range is given in each panel.
The vertical scales differ between the top and bottom panels.
}\label{fig_bin1darea_alllibvscfs}
\end{figure}

\begin{figure}
\epsscale{0.6}
\plotone{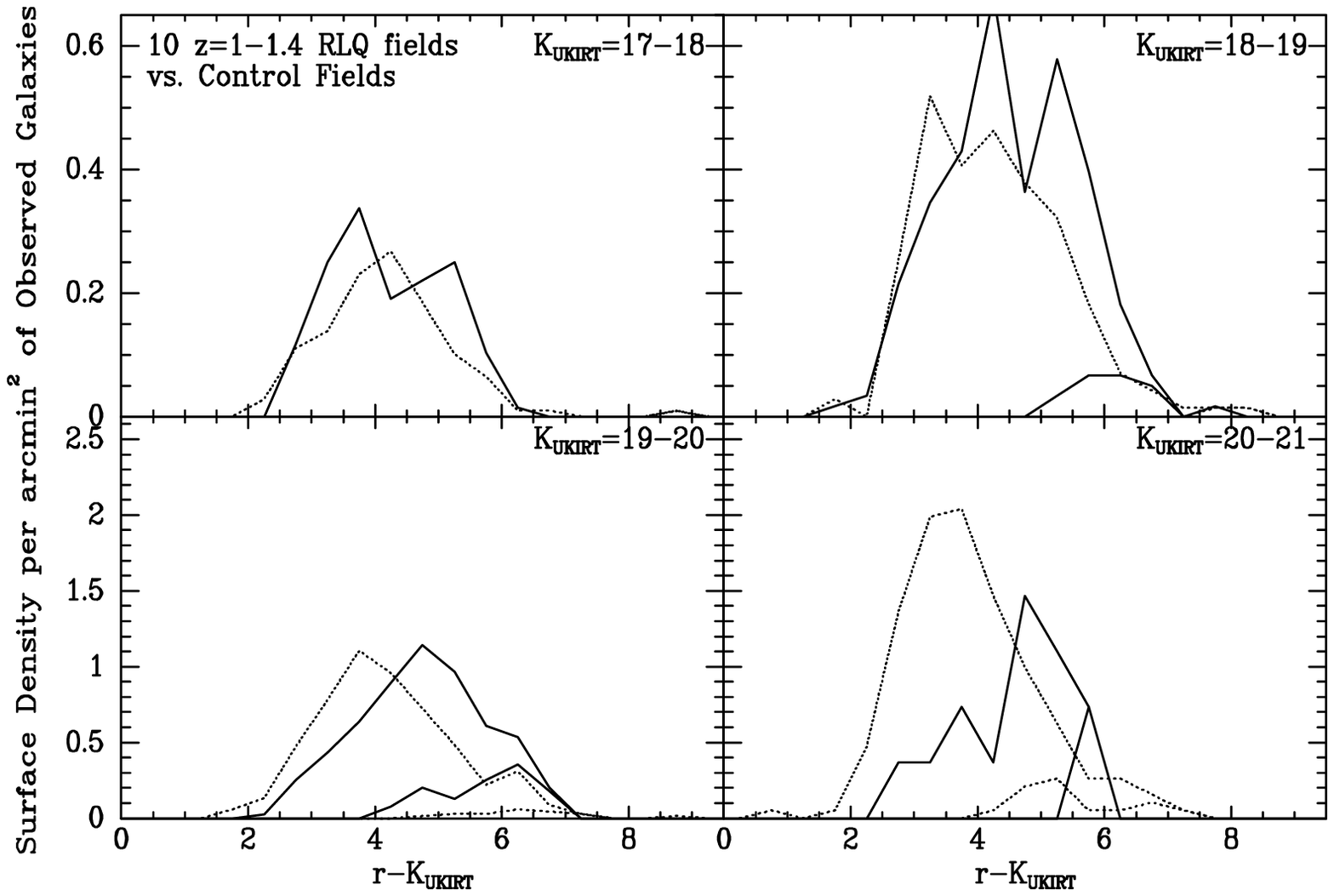}        
\caption[$r$$-$$K$ Galaxy Color Histograms in $z$=1--1.4 RLQ Fields: Surface Density of Observed Galaxies]{
\singlespace
Colors of $K$-selected galaxies.
Solid line is \z=1--1.4 quasar fields; dotted line is control fields.  Smaller
histograms represent those galaxies with lower or upper limits to their colors.
Each histogram has been normalized by the area imaged in the appropriate
magnitude range to yield the surface density of galaxies per arcmin$^2$
in each color bin.  The \Kuk\ magnitude range is given in each panel.
The vertical scales differ between the top and bottom panels.
Only a handful of galaxies in one quasar field contribute to the \Kuk=20--21
magnitude range, so the difference in those two histograms is not significant.
}\label{fig_bin1darea_midzlibvscfs}
\end{figure}

\begin{figure}
\plotone{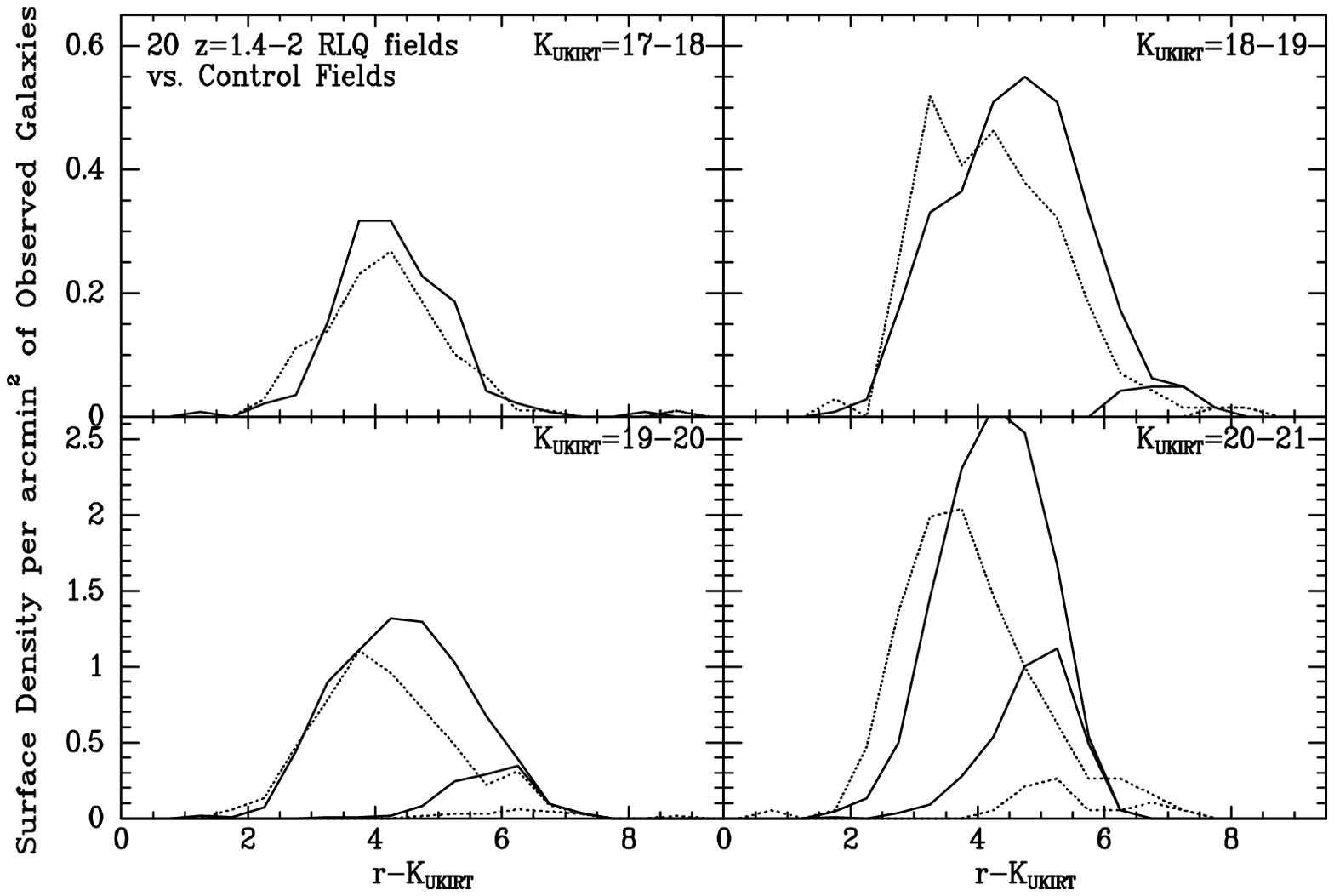}         
\caption[$r$$-$$K$ Galaxy Color Histograms in $z$=1.4--2 RLQ Fields: Surface Density of Observed Galaxies]{
\singlespace
Colors of $K$-selected galaxies.
Solid line is \z=1.4--2 quasar fields; dotted line is control fields.  Smaller
histograms represent those galaxies with lower or upper limits to their colors.
Each histogram has been normalized by the area imaged in the appropriate
magnitude range to yield the surface density of galaxies per arcmin$^2$
in each color bin.  The \Kuk\ magnitude range is given in each panel.
The vertical scales differ between the top and bottom panels.
}\label{fig_bin1darea_hizlibvscfs}
\end{figure}

\begin{figure}
\epsscale{1.0}
\plotone{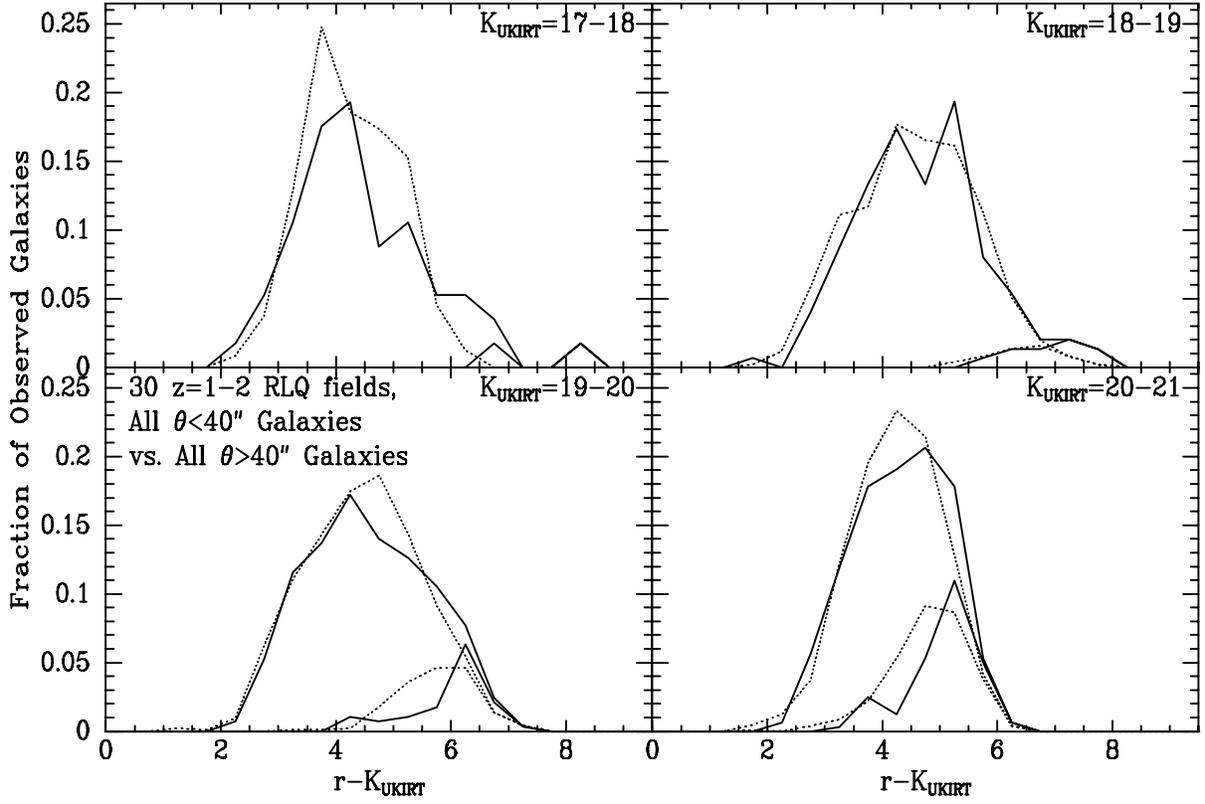} 
\caption[$r$$-$$K$ Galaxy Color Histograms in $z$=1--2 RLQ Fields: Fraction of Observed Galaxies at $\theta$$<$40$''$ and $\theta$$>$40$''$]{
\singlespace
Colors of $K$-selected galaxies in \z=1--2 RLQ fields.
Solid line is galaxies at $\theta$$<$40$''$ from the quasars;
dotted line is galaxies at $\theta$$>$40$''$ from the quasars.  Smaller
histograms represent galaxies with lower or upper limits to their colors.
The \Kuk\ magnitude range is given in each panel.
}\label{fig_bin1dnorm_alllt40libvsgt40}
\end{figure}

\clearpage

\begin{figure}
\epsscale{0.8}
\plotone{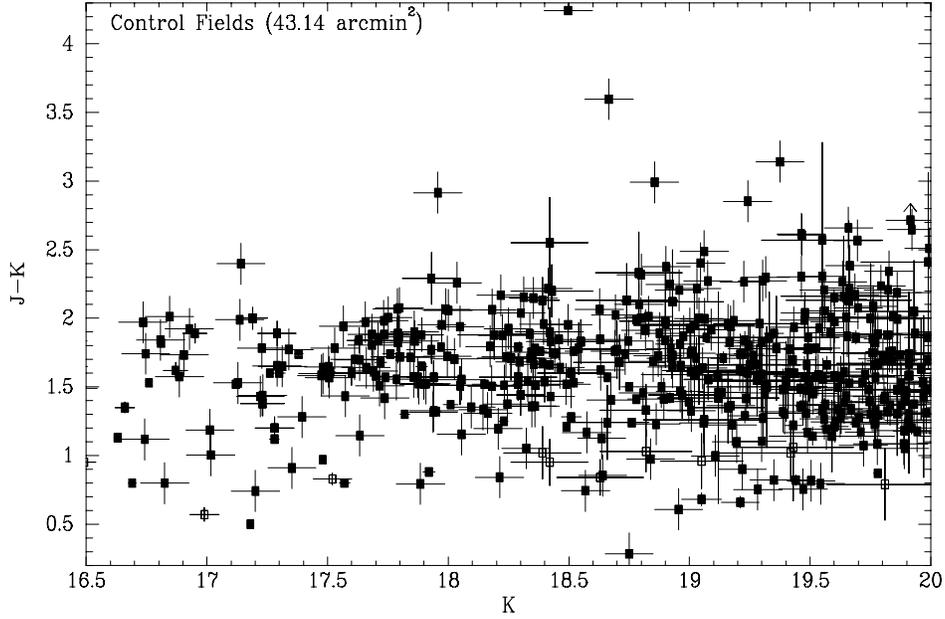}  
\caption[$K$/$J$$-$$K$ Color-Magnitude Diagram for Control Fields]{
\singlespace
$K$/$J$$-$$K$ color-magnitude diagram for galaxies in the control field
datasets of McLeod \etal\ (1995), Elston, Eisenhardt, \& Stanford (1998),
and Dickinson \etal\ (1998).  Open symbols are stars and filled symbols are 
galaxies, but classifications are only available for the McLeod dataset.
}\label{fig_kjk_cfs}
\end{figure}

\begin{figure}
\plotone{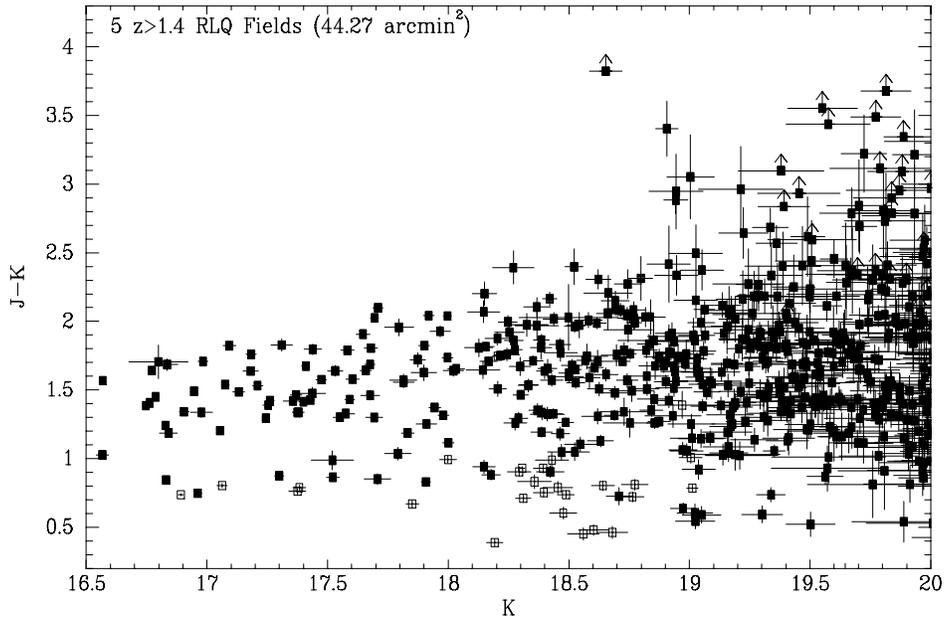}  
\caption[$K$/$J$$-$$K$ Color-Magnitude Diagram for Five $z$=1.4--2 RLQ Fields]{
\singlespace
$K$/$J$$-$$K$ color-magnitude diagram for galaxies in the five $z$=1.4--2 RLQ
fields with $J$ data.  Open symbols are stars and filled symbols are galaxies.
}\label{fig_kjk}
\end{figure}

\begin{figure}
\epsscale{0.7}
\plotone{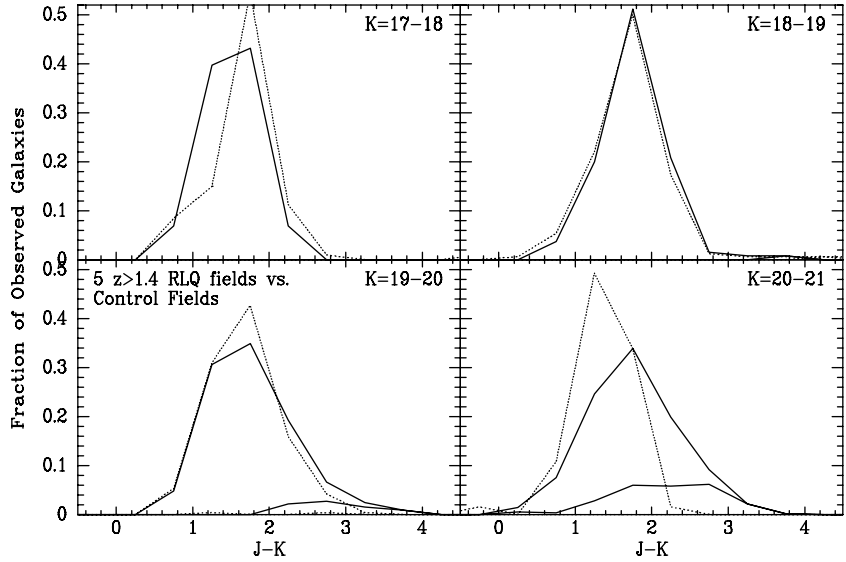}         
\caption[$J$$-$$K$ Galaxy Color Histograms in $z$=1.4--2 RLQ Fields: Fraction of Observed Galaxies]{
\singlespace
\jk\ colors of $K$-selected galaxies.
Solid line is 5 \z=1.4--2 RLQ fields with $J$ data; dotted line is control 
fields.  Smaller histograms represent those galaxies with lower or upper limits
to their colors.  Each histogram has been separated normalized to unity sum.
The \Kuk\ magnitude range is given in each panel.
}\label{fig_bin1dnorm_kjkvscfs}
\end{figure}

\begin{figure}
\plotone{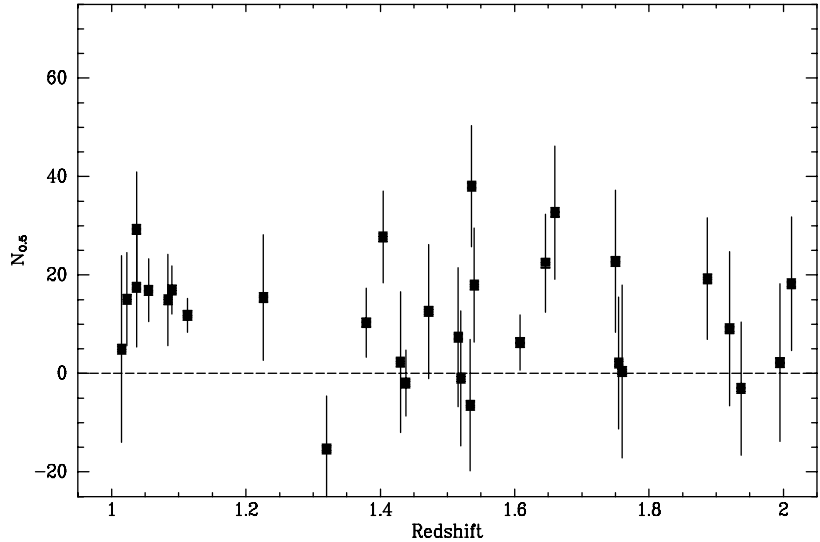}		
\caption[``Near-Field'' Richness $N_{0.5}$ vs. Redshift]{
\singlespace
The Hill \& Lilly (1991) richness statistic \no5\ measured in our RLQ fields.
Error bars are Poisson only.
The horizontal dashed line at \no5=0 shows where the observed counts at 
$<$0.5~Mpc equal the prediction from data at $>$0.5~Mpc.
}\label{fig_n05adj}
\end{figure}

\clearpage

\begin{figure}
\plotone{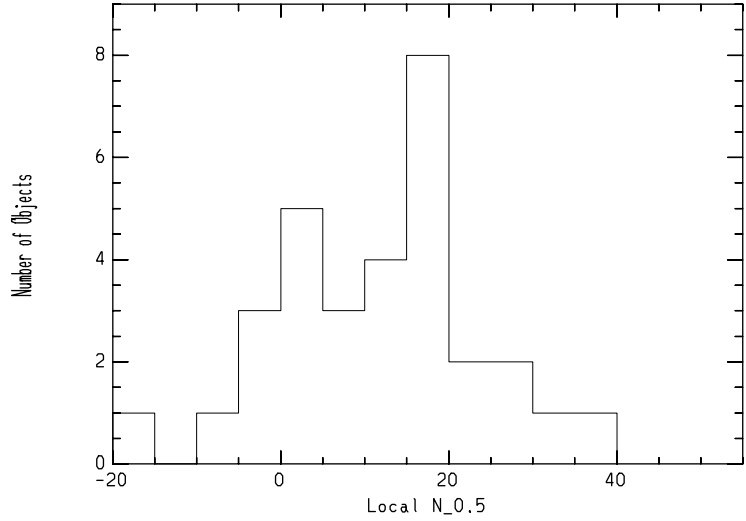}		
\caption[Histogram of ``Near-Field'' Richness $N_{0.5}$ Values]{
\singlespace
Histogram of ``near-field'' \no5\ values measured for our RLQ fields.
Abell richnesses of 0, 1, and 2 are approximately \no5=10, 19, and 31.
}\label{fig_n05adjhist}
\end{figure}

\begin{figure}
\plotone{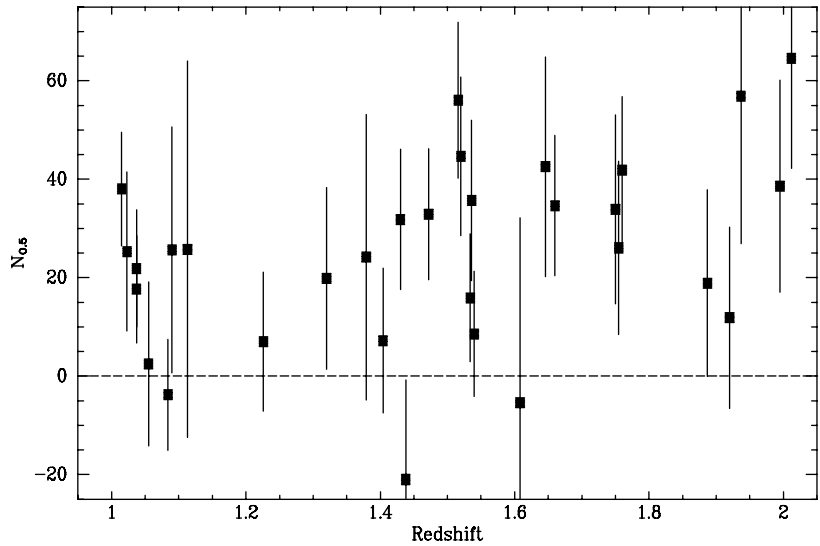}		
\caption[``Far-Field'' Richness $N_{0.5}$ vs. Redshift]{
\singlespace
The Hill \& Lilly (1991) richness statistic \no5\ measured for our RLQ fields
compared to the literature.  Error bars include the uncertainties in the
corrections for fields which do not reach $K_{BCG}$+3.
The horizontal dashed line at \no5=0 shows where the observed counts at 
$<$0.5~Mpc equal the prediction from the average published literature data.
}\label{fig_n05vslit}
\end{figure}

\begin{figure}  
\epsscale{0.55}
\plotone{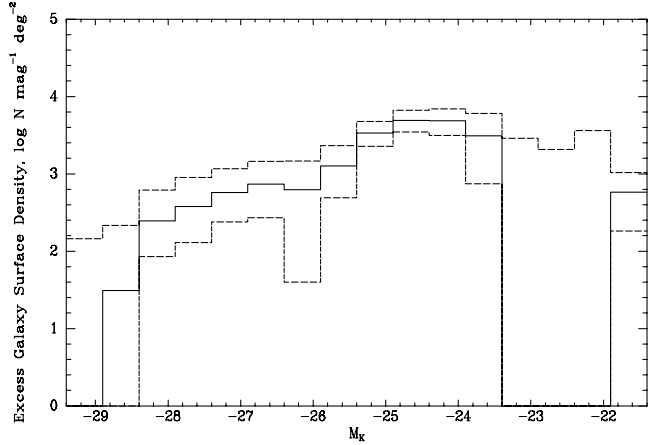}
\caption[Excess Galaxy Surface Density in \z$<$1.4 fields vs. $M_K$]{
\singlespace
The solid line is the excess galaxy surface density (log $N$) vs. $M_K$.
Dashed lines show the $\pm$1$\sigma$ uncertainty envelope.  
The excesses from all \z$<$1.4 fields have been coadded and
normalized by the area at each magnitude bin to make this plot.
}\label{fig_klfmidzm_k}
\end{figure}

\begin{figure}  
\epsscale{0.55}
\plotone{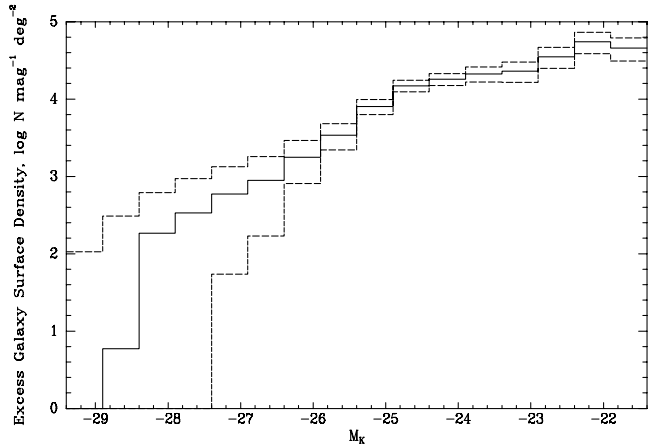}
\caption[Excess Galaxy Surface Density in \z$>$1.4 fields vs. $M_K$]{
\singlespace
The solid line is the excess galaxy surface density (log $N$) vs. $M_K$.
Dashed lines show the $\pm$1$\sigma$ uncertainty envelope.  
The excesses from all \z$>$1.4 fields have been coadded and normalized by the
area at each magnitude bin to make this plot.  Even though these fields are at
higher redshift, better data enables us to reach farther down the luminosity
function in them than in the \z$<$1.4 fields, on average.
}\label{fig_klfhizm_k}
\end{figure}

\begin{figure}
\caption[Q~0835+580 Field $rJK_s$ Color Image]{
\singlespace
Color image of the field of Q~0835+580 (\z=1.534)
using $rJK_s$ images to drive the blue, green and red color guns, respectively.
Saturation occurs at $r$=24, \ks=20, and $J$=21 mag/arcsec$^2$.  North is up
and East is left; the area in black is approximately 3.25$'$ by 3.25$'$.
}\label{fig_rjk0835}
\end{figure}

\begin{figure}
\caption[Q~1126+101 Field $rJK_s$ Color Image]{
\singlespace
Color image of the field of Q~1126+101 (\z=1.516).
See key to Figure \ref{fig_rjk0835} for details.
The area in black is approximately 3.3$'$ by 3.3$'$.
}\label{fig_rjk1126}
\end{figure}

\begin{figure} 
\epsscale{0.775}
\plotone{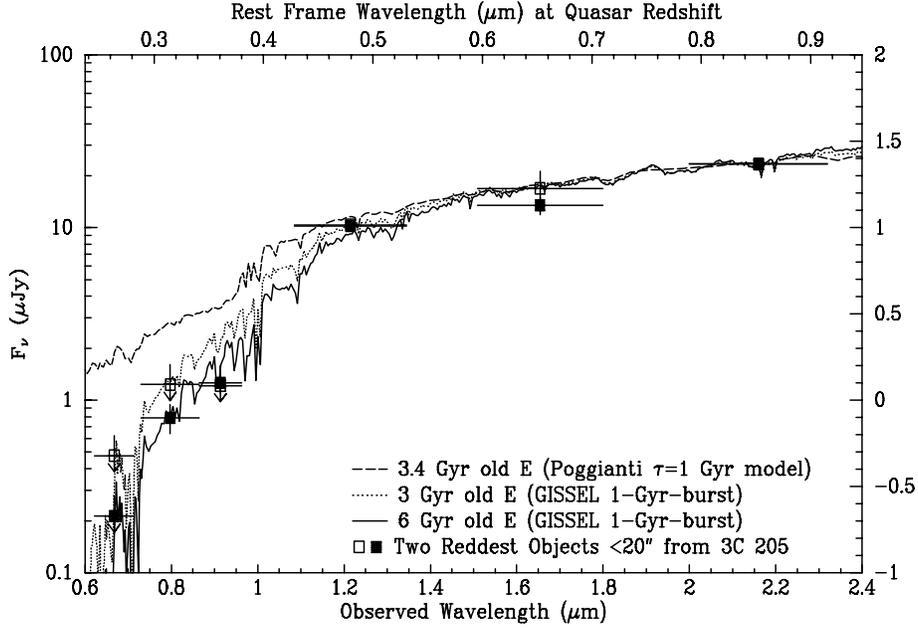}
\caption[SED and Fits for 2 Reddest Objects $<$20$''$ from 3C~205]{
\singlespace
The SEDs for the two reddest objects in \rks\ within \th=20\farcs5 of Q~0835+580
(\z=1.534) are shown as the filled and open squares with Poisson error bars
on the fluxes.  Data points are plotted at the effective wavelengths of each
filter,
and horizontal error bars indicate the widths of the filters.
Model spectra are as shown on the figure and discussed in the text.
}\label{fig_sed0835nrq2red}
\end{figure}

\begin{figure} 
\epsscale{0.775}
\plotone{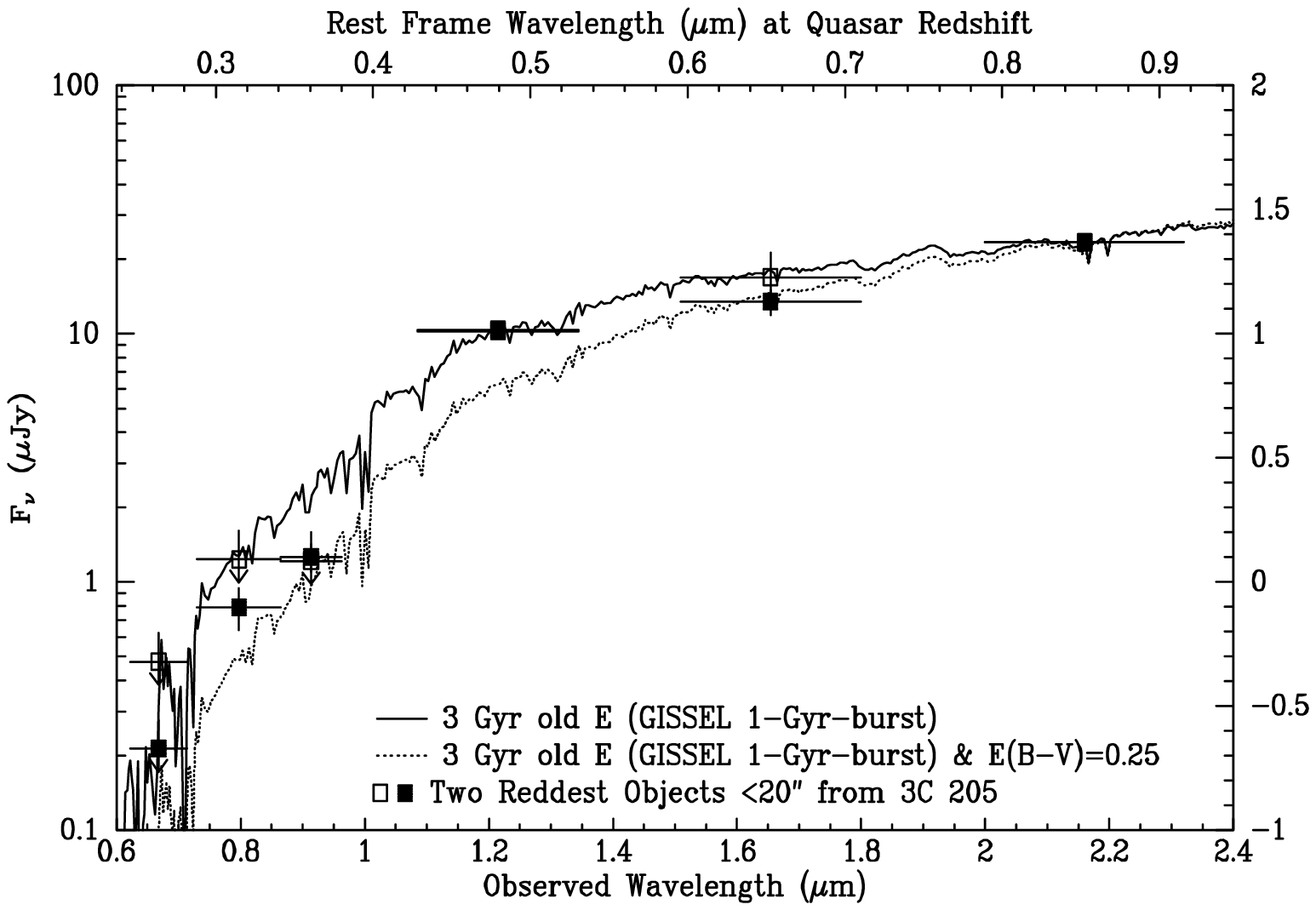}
\caption[SED and Dust-Reddened Fits for 2 Reddest Objects $<$20$''$ from 3C~205]{
\singlespace
See key to Figure~\ref{fig_sed0835nrq2red} for details.
Model spectra are as shown on the figure and discussed in the text.
}\label{fig_sed0835nrq2red_dust}
\end{figure}

\begin{figure} 
\epsscale{0.775}
\plotone{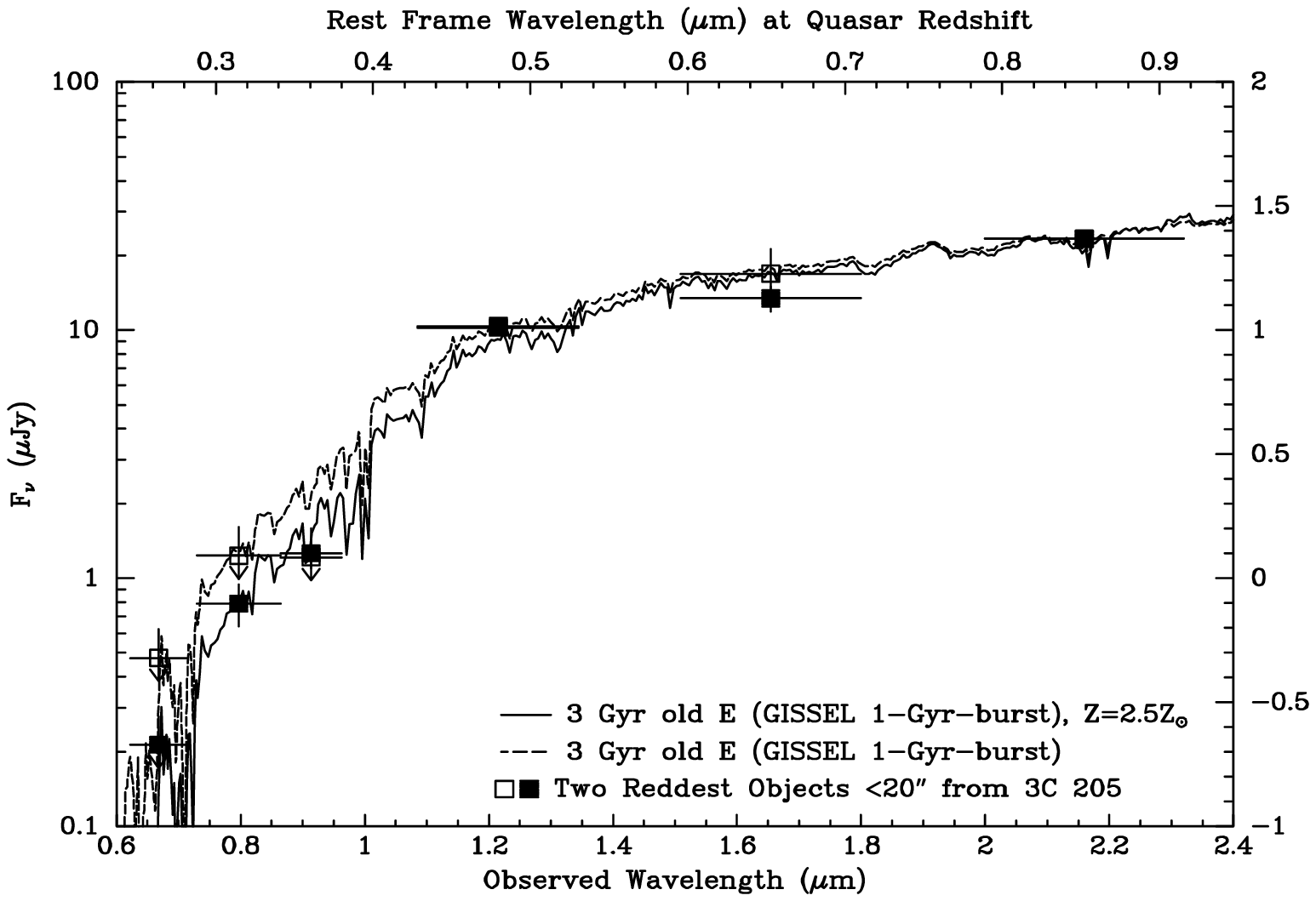}
\caption[SED and Different-Metallicity Fits for 2 Reddest Objects $<$20$''$ from 3C~205]{
\singlespace
See key to Figure~\ref{fig_sed0835nrq2red} for details.
Model spectra are as shown on the figure and discussed in the text.
}\label{fig_sed0835nrq2red_metal}
\end{figure}

\begin{figure} 
\epsscale{0.775}
\plotone{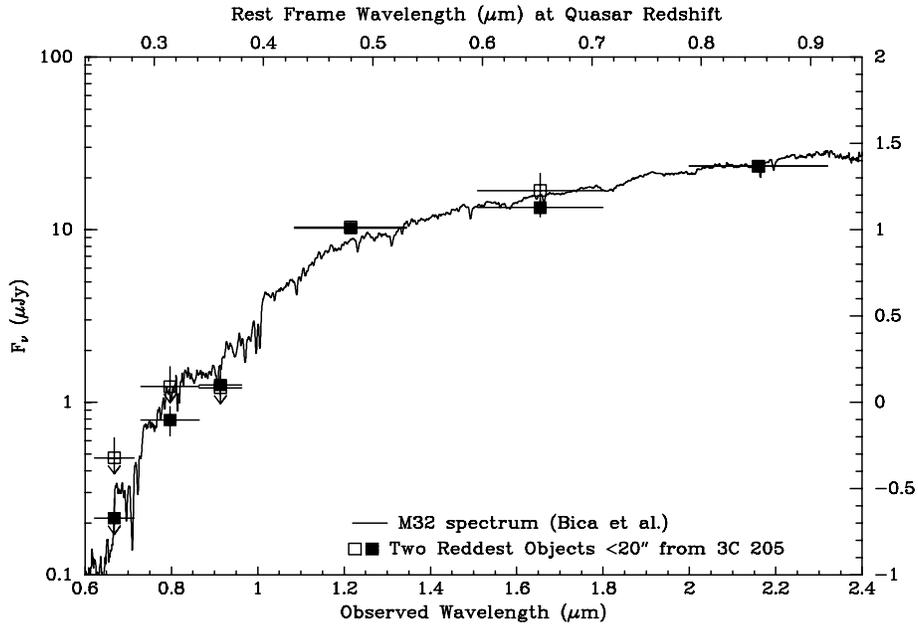}
\caption[SED for M32 and 2 Reddest Objects $<$20$''$ from 3C~205]{
\singlespace
The SEDs for the two reddest objects in \rks\ within \th=20\farcs5 of Q~0835+580
(\z=1.534) are shown as the filled and open squares with Poisson error bars
on the fluxes.  Horizontal error bars indicate the widths of the filters used
to construct the SEDs.  The Bica \etal\ (1996) spectrum of M32 is shown as the
solid line.
}\label{fig_sed0835nrq2redvsm32}
\end{figure}

\begin{figure} 
\epsscale{1.0}
\plotone{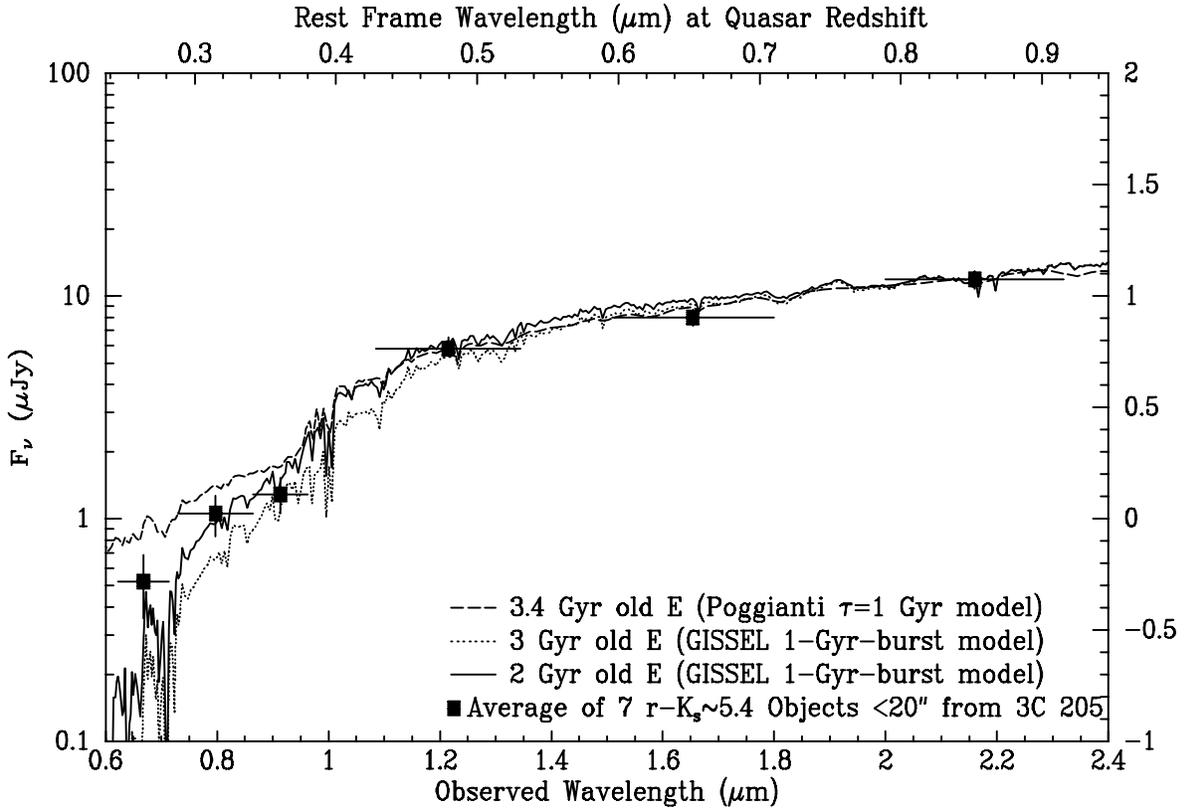}
\caption[SED and Fits for 7 Red Objects $<$20$''$ from 3C~205]{
\singlespace
The average SED for the seven next reddest objects in \rks\ within
\th=20\farcs5 from Q~0835+580 (\z=1.534) is shown as the filled squares and 
Poisson error bars on the fluxes.  Horizontal error bars indicate the widths of
the filters used to construct the SEDs.  Model spectra are as shown on the
figure and discussed in the text.
}\label{fig_sed0835nrq7red}
\end{figure}

\begin{figure} 
\epsscale{1.0}
\plotone{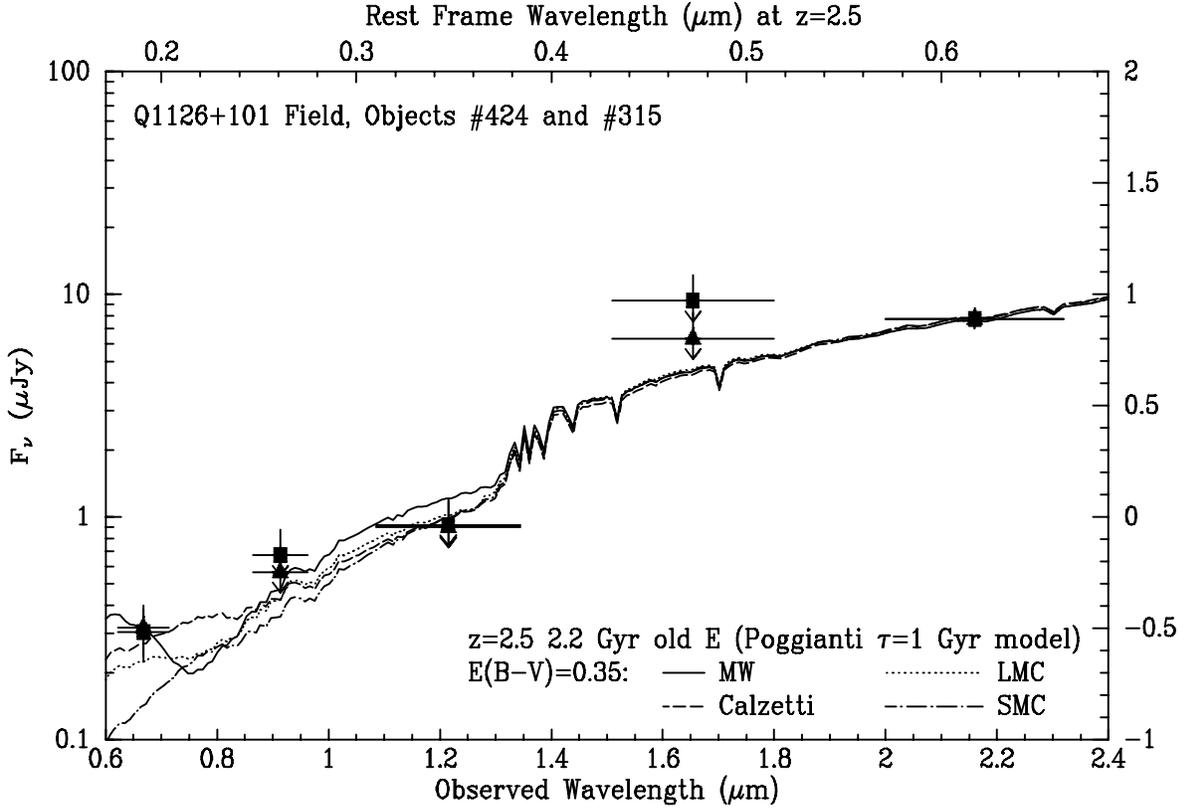}
\caption[SEDs of Two $z$$>$2 Candidates Compared to Dust-Reddened Spectra]{
\singlespace
The data points are the two objects with \rks$>$5 which are reddest in \jks\ in
the Q~1126+101 field,
\#424 (triangles) and \#315 (squares), normalized to the same flux in \ks.
The lines are the 2.2~Gyr old E model of Poggianti (1997) reddened by different
extinction laws with \ebv=0.35 and normalized to the data in \ks.
Solid line is the Milky Way (MW) extinction law from Seaton (1979) and Howarth
(1983).  Dashed line is the extinction law from Calzetti (1997).  Dotted line
is the LMC extinction law from Howarth (1983).  Dash-dot line is the SMC
extinction law given by a linear (in 1/$\lambda$) fit to the data of
Pr\'evot \etal\ (1984).  All extinction laws have been normalized to $R_V$=3.
The relative extinctions for the various laws are significantly different only
at rest-frame $\lambda$$<$3800~\AA.
}\label{fig_sedust1126b}
\end{figure}

\begin{figure} 
\epsscale{0.65}
\plotone{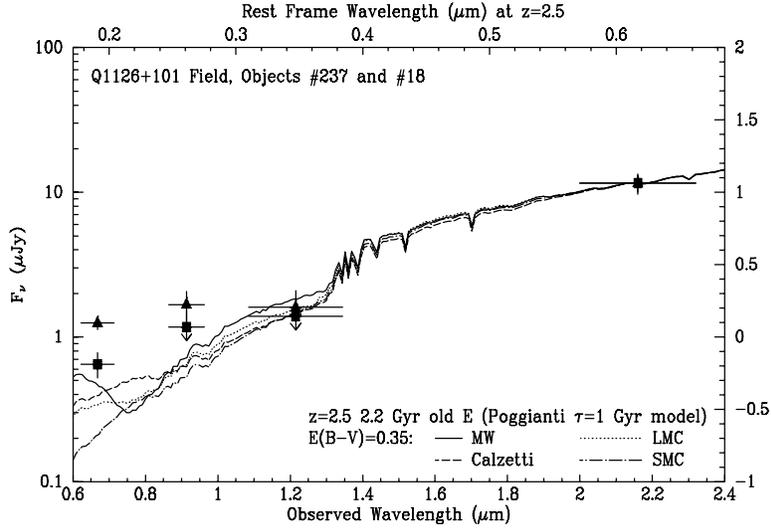}
\caption[SEDs of Another Two $z$$>$2 Candidates Compared to Dust-Reddened Spectra]{
\singlespace
The data points are the two objects with \rks$<$5 which are reddest in \jks\ in
the Q~1126+101 field, \#237 (triangles) and \#18 (squares), normalized to the 
same flux in \ks.  The lines are the 2.2~Gyr old E model of Poggianti (1997)
reddened by the same extinction laws as in Figure~\ref{fig_sedust1126b}, with
\ebv=0.35.
}\label{fig_sedust1126c}
\end{figure}

\begin{figure} 
\plotone{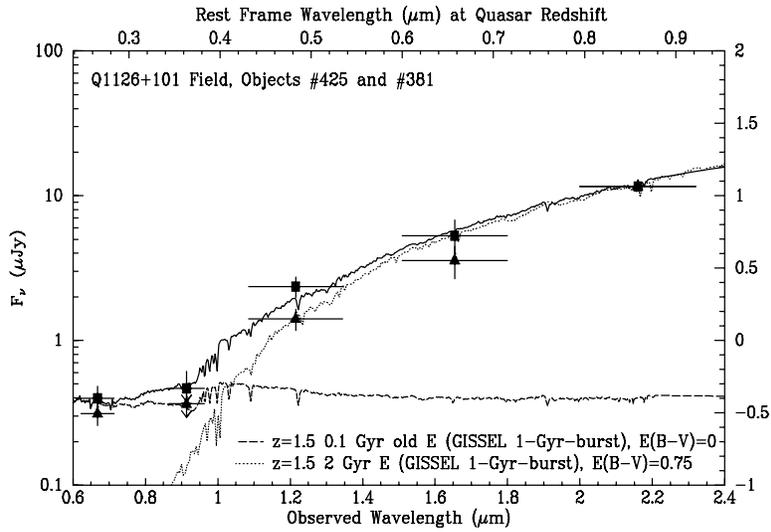}
\caption[SEDs of Another Two $z$$>$2 Candidates Compared to Dust-Reddened Spectra]{
\singlespace
The data points are the two objects with \jks$>$2.5 in the Q~1126+101 field
which have believable detections in $H$, \#425 (triangles) and \#381 (squares),
normalized to the same flux in \ks.  The dotted line is a 2~Gyr old GISSEL
model E reddened by \ebv=0.75 using the extinction law from Calzetti (1997)
and normalized to the data in \ks.  The dashed line is a 0.1~Gyr old unreddened
GISSEL model E normalized to the data in $r$.  The solid line is the sum of
the two.
}\label{fig_sedust1126d}
\end{figure}

\end{document}